\documentclass[pdflatex,sn-mathphys-num]{sn-jnl}

\usepackage{graphicx}
\usepackage{multirow}
\usepackage{amsmath,amssymb,amsfonts}
\usepackage{amsthm}
\usepackage{mathrsfs}
\usepackage[title]{appendix}
\usepackage{xcolor}
\usepackage{textcomp}
\usepackage{manyfoot}
\usepackage{booktabs}
\usepackage{algorithm}
\usepackage{algorithmicx}
\usepackage{algpseudocode}
\usepackage{listings}
\usepackage{stmaryrd}

\def\sech{{\rm sech}}
\def\cn{{\rm cn}}

\begin{document}

\title[On shallow water non-convex dispersive hydrodynamics: the extended KdV model]
{On shallow water non-convex dispersive hydrodynamics: the extended KdV model}

\author*[1]{\fnm{Saleh} \sur{Baqer}}
\email{saleh.baqer@ku.edu.kw}

\author[2]{\fnm{Theodoros P.} \sur{Horikis}}
\email{horikis@uoi.gr}

\author[3]{\fnm{Dimitrios J.} \sur{Frantzeskakis}}
\email{dfrantz@phys.uoa.gr}

\affil[1]{Department of Mathematics, Kuwait University, Kuwait City 13060}
\affil[2]{Department of Mathematics, University of Ioannina, Ioannina 45110, Greece}
\affil[3]{Department of Physics, National and Kapodistrian University of Athens, Athens 15784, Greece}


\abstract{
In this work, we investigate non-classical wavetrain formations, and particularly dispersive shock waves (DSWs), or undular bores, in systems exhibiting non-convex dispersion. Our prototypical model, which arises in shallow water wave theory, is the extended Korteweg-de Vries (eKdV) equation. The higher-order dispersive and nonlinear terms of the latter, lead to resonance between dispersive radiation and solitary waves, and notably, the individual waves comprising DSWs, due to non-convex dispersion. This resonance manifests as a resonant wavetrain propagating ahead of the dispersive shock wave.
We present a succinct overview of the fundamental principles and characteristics of DSWs and explore analytical methods for their analysis. Wherever applicable, we demonstrate these concepts and techniques using both the classical KdV equation and its higher-order eKdV counterpart. We extend the application of the dispersive shock fitting method and the equal amplitude approximation to investigate radiating DSWs governed by the eKdV equation. We also introduce Whitham shock solutions for the regime of traveling DSWs of the eKdV model. Theoretical predictions are subsequently validated against direct numerical solutions, revealing a high degree of agreement.}


\maketitle

\section{Introduction}

The theory of the water waves \cite{stoker,johnson,lannes}, which is being developing 
for more than two centuries, plays a particularly important role in applied 
mathematics and physics. Indeed, while the water wave problem is extremely 
interesting and important on its own right, it has provided the background
for the development of the theory of nonlinear dispersive waves, which has 
been a key element in numerous disciplines \cite{dodd,infeld,mja}.
 
Generally, since water waves propagate on a free surface, i.e., the surface itself is 
not fixed but evolves dynamically, the study of water waves involves a nonlinear free 
boundary value problem, which demands sophisticated mathematical tools for their 
analysis \cite{stoker,johnson,lannes,dodd,infeld,mja,whitham}. 
The shallow water wave problem is a particular scenario focusing on the behavior 
of surface waves in regions where the depth of the fluid is much smaller compared 
to the wavelength of the waves. Studies on shallow water waves commonly rely on 
physically relevant models describing ideal fluids, which are assumed to be 
incompressible, inviscid, and irrotational. Under these assumptions, and using 
asymptotic expansion methods, a variety of effective shallow water wave equations 
can be derived from the Euler (or water wave) equations \cite{lannesnonlin,us}. 
Out of these models, the Korteweg-de Vries (KdV) equation stands as a fundamental 
model, offering valuable insights into the behavior and dynamics of water waves. 

The KdV equation is often referred to as a ``{\em universal equation}'' because 
of its remarkable ability to describe a wide range of nonlinear wave phenomena across 
various fields \cite{dodd,infeld}. One reason for the KdV equation's universality lies 
in its balance between nonlinear and dispersive effects, which allows for an  
accurate description of fundamental nonlinear waveforms, such as solitons  
and dispersive shock waves (DSWs) ---{\it alias} undular bores in hydrodynamics
\cite{gennel}. DSWs arise from the nonlinearity-induced steepening of wave 
fronts in weakly dispersive media, and manifest as highly nonlinear wave packets 
characterized by a leading shock front followed by dispersive oscillations. 
DSWs are common not only in hydrodynamics, but also in a variety of physical 
contexts, including meteorology \cite{loweratmosphere,morning1,morning2}, 
plasma physics\cite{plasma1,plasma2,plasma3,plasma4,plasmapit,shockyannis1,
shockyannis2,shockmagneto}, traffic flow \cite{lighthilltraffic,benjamintraffic1, benjamintraffic2, benjamintraffic3}, photorefractive crystals 
\cite{fleischer1,fleischer2,elopt,wan}, optical fibers \cite{trillohump,fatome}, 
thermal optical media \cite{trilloshocks,wang}, nematic liquid crystals 
\cite{nemboreori,nemboreel,salehnem1,salehnem2}, quantum fluids including 
Bose-Einstein condensates \cite{bose1,bose2,bose3,bose4,dswdroplet}, 
solid mechanics \cite{karimasolidbore1,karimasolidbore2,karimasi, purohit}, 
nonlinear dynamical lattices \cite{latticebore,patdiscdsw}, and erosion modeling \cite{erosion}; 
see also the review~\cite{elreview} and 
references therein.

In what follows, we give a concise overview of the fundamental principles and 
characteristics governing DSWs, and explore recent advancements in the emerging field 
of {\it non-convex dispersive hydrodynamics}. The latter, is a branch of nonlinear 
dispersive wave theory that focuses on the investigation of non-classical wavetrain  
formations, such as dispersive shock waves (and solitary waves) in systems exhibiting 
non-convex dispersion (see, e.g., Ref.~\cite{sande}). 
Throughout, we illustrate these concepts and techniques using both the classical KdV 
equation and its higher-order counterpart, the extended KdV (eKdV) equation. This  
model is a higher-order nonlinear dispersive wave equation, derived from the Euler 
equations when the asymptotic expansion is taken to one order beyond the KdV 
approximation~\cite{us,ekdv,ekdhighermodu,salehekdv}. 
To be more specific, a brief description of our presentation, and the organization of 
this work are as follows. 

First, in Section~2, we introduce basic notions of non-convex hydrodynamics and 
shallow water DSWs in the context of the eKdV equation. Then, we show that the 
presence of the next higher order dispersive, nonlinear, and nonlinear-dispersive 
terms in the latter, lead to resonance between dispersive radiation and solitary 
waves, and notably, the individual waves comprising DSWs, due to non-convex 
dispersion. This resonance manifests as a resonant wavetrain propagating ahead of the 
dispersive shock wave. Non-convex DSW regimes, namely radiating DSW (RDSW), cross-over 
DSW (CDSW), and traveling DSW (TDSW) are identified and illustrated. 

Then, in Section~3, we present an overview of the mathematical tools that are used 
for the analysis of non-convex DSWs. We start with the Whitham modulation theory, 
and present both approaches, namely the one involving the averaging of conservation 
laws \cite{whithampert}, and the other employing the averaging of Lagrangians 
\cite{whithamvar1,whithamvar2}. Then, we discuss the dispersive shock fitting method, 
devised by El \cite{fitting}, which allows for the determination of DSW's macroscopic 
properties (such as the velocities of the leading solitary wave edge and of the 
trailing harmonic edge), relying solely on the associated \textit{linear} dispersion 
relation. 
Furthermore, we examine the admissibility conditions \cite{hoefersci}, which 
are necessary to maintain the stable form of eKdV shallow water DSWs, and discuss the 
so-called equal amplitude approximation method \cite{equalamp}, which enables the 
determination of certain properties of unstable DSWs (e.g., the amplitude and velocity 
of the leading solitary wave edge, and the number of solitary waves at a given time). 
We apply these methods to investigate radiating undular bores governed by the eKdV equation with shallow water wave coefficients, comparing their predictions with those obtained from higher-order modulation theory. 
In addition, in the same Section (Sec.~3), we introduce Whitham shock 
solutions --first introduced by Sprenger and Hoefer \cite{patjump}-- 
for the regime of the TDSWs of the eKdV equation. It is found that this 
regime constitutes a special case of the regime of CDSWs of the eKdV, occurring when 
the amplitude of the lead solitary wave of the DSW diminishes.

In Section~4, we corroborate our analytical predictions by presenting 
results of direct numerical simulations; the agreement between the analytical and 
numerical results is excellent. Finally, in Section~5, we present our conclusions and 
discuss interesting directions for future studies.   

\section{Shallow water DSWs and non-convex effects}  

Broadly speaking, solutions of nonlinear hyperbolic partial differential equations  
(PDEs), supplemented even by smooth 
initial data, may develop singularities in finite time, that is, shock waves. 
In fact, these structures, which emerge when there is rapid variation in physical  
quantities due to nonlinearity \cite{whitham}, arise in diverse physical 
applications, involving dissipative and dispersive media. For instance, in gas 
dynamics, a shock wave forms when an evolving physical quantity, like fluid velocity, 
temperature, pressure, or density, exceeds the speed of sound. This results in 
supersonic flow with a Mach number ${\rm M}>1$ and leads to a sonic boom. On the other 
hand, in shallow water wave theory, a shock wave emerges when fluid moves at a 
velocity surpassing the linear shallow water wave velocity, $\sqrt{gh}$ (where $g$ is 
the  acceleration due to gravity and $h$ is the water depth). This condition 
corresponds to a Froude number ${\rm Fr}>1$, indicating a supercritical flow. 

The most comprehensive and simplest scalar, one-dimensional mathematical model featuring a shock wave solution is the nonlinear hyperbolic PDE:
\begin{equation}
u_{t} + f_{x}(u) = 0,\label{e:gmodel}
\end{equation}
where $u$ is the density and $f$ is the flux function. For generating a right-propagating shock, the simplest initial condition is the step initial condition 
\begin{equation}\label{ic_shock}
    u(x,0)= \left\{ \begin{array}{cc}
         u_{-},\quad x<x_{0},\\
         u_{+},\quad x>x_{0}, 
    \end{array}\right. 
\end{equation}
with the assumption that $u_{-}>u_{+}$ and $f^{\prime}(u)>0$. The step initial condition promptly breaks as the time evolves, leading to a shock formation at $x=x_{0}$. This initial value problem is termed a Riemann problem. The flux function $f$ is a smooth function with $f^{\prime\prime}(u)\neq{0}$, which is a genuine nonlinearity condition in hydrodynamics. If $f^{\prime}(u)<0$, a shock wave forms only when $u_{-}<u_{+}$; otherwise, a left-propagating rarefaction wave occurs. The case of $u_{-}<u_{+}$ with $f^{\prime}(u)>0$ leads to a right-propagating rarefaction wave \cite{whitham,kamchatnovbook}. The formation of shocks becomes more intricate when $f^{\prime}(u)>0$ for some values of $u$ and $f^{\prime}(u)<0$ for others. In such scenarios, multiple shock waves traveling in different directions can be generated --see for instance Ref.~\cite{helium}. 

In a Riemann problem involving shock waves, the immediate breakdown of the initial 
condition results in the generation of multiple distinct values of a physical quantity 
at a single spatial position, due to the effect of nonlinearity. This can manifest as 
different fluid density or velocity values at a given one position $x$, which is 
evidently non-physical. Mathematically speaking, this breaking process leads to a 
blow-up in the derivative values, and this is termed a gradient catastrophe 
\cite{whitham,elreview}. The resolution of this singularity, namely the 
regularization of the discontinuity, hinges on the intrinsic nature of the physical 
medium, whether it exhibits viscosity or dispersion effects. In the present work, we 
focus on the latter case. Examples of dispersive media include shallow waters, 
quantum fluids, fiber optics, thermal optical materials, and nematic liquid crystals, 
as mentioned in the previous Section. To rectify the physical singularity, the 
nonlinear PDE (\ref{e:gmodel}) requires a correction through the incorporation of a 
non-zero differential operator or integro-differential operator $D[u]$, namely, 
\begin{equation}
u_{t} + f_{x}(u) = D[u], 
\label{nldm}
\end{equation}
where $D[u]$ should contain spatial or mixed higher-order derivatives, resulting 
in a \textit{real-valued} linear dispersion relation.

\begin{figure}
    \centering
\includegraphics[width=1.0\textwidth]{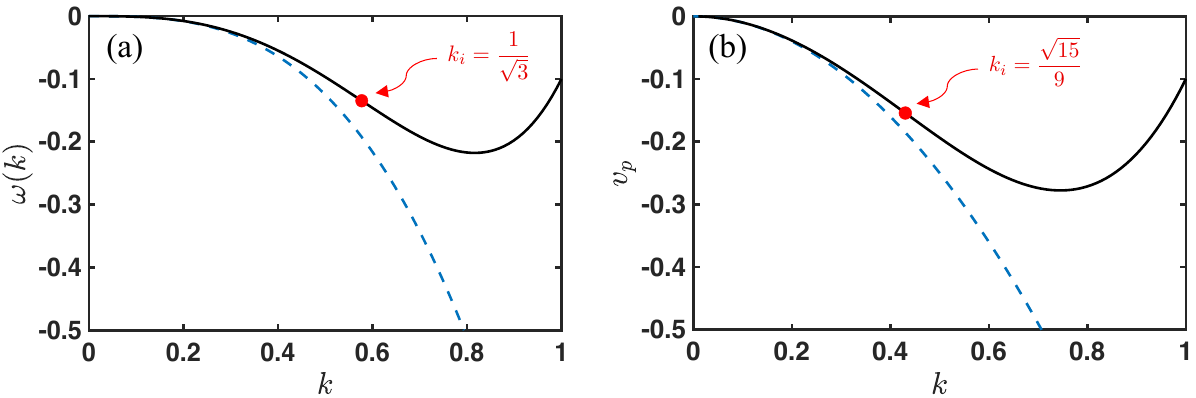}
    \caption{KdV and eKdV linear dispersion relation curves (a) and phase velocity curves (b), demonstrating the non-convexity effect arising from higher order terms. The solid black lines correspond to the eKdV equation (\ref{e:ekdv}), whereas the red dashed lines correspond to the KdV equation (\ref{e:kdv}). The red point on the curve shows the location of the zero dispersion point $k_i$, i.e., the inflection point where $\omega_{kk}=0$. Here, $\bar{u}=0$, $c_1=-1$, $c_2=1$, $c_3=1$, $c_4=6.0$ and $\epsilon=0.15$. (Color version online).}
    \label{f:dispersion}
\end{figure}

The KdV equation, a classical model frequently employed as a benchmark in dispersive hydrodynamics, effectively captures the influence of dispersive effects on shock waves in media with dispersion.  
In dimensionless form, the KdV equation reads:
\begin{equation}
u_{t}+6uu_{x}+u_{xxx}=0,
\label{e:kdv}
\end{equation}
where $u(x,t)$ denotes the water wave elevation from an equilibrium state. The linear dispersion relation of this equation on the background $\bar{u}$ is concave: 
$\omega(k;\bar{u})=6\bar{u}k-k^{3}$, with $0<k<1$ (long-wave limit). Another 
pertinent yet non-trivial dispersive hydrodynamic model is an extension of the KdV equation, encompassing higher-order nonlinear, dispersive and nonlinear-dispersive terms, which is referred to as the extended KdV (eKdV) equation 
\cite{us,ekdv,ekdhighermodu,salehekdv}:
\begin{equation}
u_{t} + 6uu_{x} + u_{xxx} + \epsilon \left( c_{1} u^{2}u_{x} + c_{2} u_{x}u_{xx} + c_{3} uu_{xxx} + c_{4} u_{xxxxx} \right) = 0. 
\label{e:ekdv}
\end{equation}
Here, the parameter $\epsilon$ 
quantifies the strength of the dispersion effect or weak nonlinearity, 
representing, for water waves, 
the wave amplitude to the undisturbed fluid depth ratio. In the specific context of shallow water waves, the higher-order coefficients take the values 
(see, e.g., Ref.~\cite{salehekdv}):
\begin{equation}
    c_{1} = -3/2,\,c_{2} = 23/4,\,c_{3} = 5/2,\,\text{and}\,c_{4} = 19/40.\label{e:shallowcoef}
\end{equation}
The eKdV equation finds applications in the study of gravity-capillary waves for 
which the Bond number is near 1/3. Specifically, the equation in this case corresponds to the Kawahara equation with $c_{1}=c_{2}=c_{3}=0$ and $c_{4}\neq{0}$ \cite{kawahara,patkawahara,patkawtrav}, 
\begin{equation}
u_{t} + 6uu_{x} + u_{xxx} + \epsilon c_{4} u_{xxxxx} = 0. \label{e:kawahara}
\end{equation}
The associated linear dispersion on the background $\bar{u}$ in this setting, 
\begin{equation}
    \omega(k;\bar{u})= (6\bar{u}+\epsilon c_{1}\bar{u}^{2})k-\left(1+\epsilon c_{3}\bar{u}\right)k^3 + \epsilon c_{4} k^5, 
    \label{e:omegaekdv}
\end{equation}
can exhibit non-convex behavior for particular coefficients. 
Indeed, in Fig.~\ref{f:dispersion} we illustrate how the inclusion of higher-order terms results in non-convexity in the
linear dispersion relations, as well as in the phase velocity profiles, 
for both the KdV and eKdV equations. 
The non-convex nature of dispersion is a pivotal characteristic in dispersive hydrodynamic systems, exerting a profound influence on the structures of governed dispersive shock waves. We will discuss these effects in detail below. 

In dissipative media, exemplified by the flow of a compressible gas modelled 
mathematically by the Burgers equation ($f(u)=u^2/2$ and $D[u]=\nu u_{xx}$ in 
Eq.~(\ref{nldm}), with $\nu$ being the viscosity strength), the dominance of 
viscosity over dispersion leads to the appearance of a hyperbolic tangent front 
as the resolution for the physical singularity. This smooth front serves to connect 
the steady initial states $u_{-}$ and $u_{+}$. In contrast, in dispersive media, where 
the effect of dispersion takes over dissipation ---or dispersion is non-existent in 
the first place--- the wave breaking is resolved by a non-steady (i.e., spatially 
expanding) slowly-varying wavetrain that connects the stationary initial levels 
$u_{-}$ and $u_{+}$. This wavetrain is referred to as a dispersive shock wave (DSW) or, as frequently termed in fluid mechanics, undular bore. 

DSWs are inherently multi-scale wavetrains, composed of two distinct scales. 
They feature fast-scale wave parameters, namely, the wave phase $\theta(x,t)$, the 
wavelength $L(x,t)$, and the period $T(x,t)$, alongside slow-scale wave parameters, 
encompassing the amplitude $a(x,t)$, the frequency $\omega(x,t)$, the wavenumber 
$k(x,t)$, and the mean level $\bar{u}(x,t)$ connecting the initial level behind 
$u_{-}$ with the initial level ahead $u_{+}$. The nonlinearity in DSW 
propagation is notably robust, as the leading and trailing edges propagate with 
markedly different velocities. The leading edge is characterized by a solitary wave 
and travels at the solitonic velocity $s_{+}$. In contrast, the trailing edge is 
defined by a train of harmonic waves that propagate with the group velocity 
$s_{-}=\omega_{k}(k_{-};u_{-})$. Macroscopic and microscopic properties 
of DSWs, including the edge velocities, can be determined using Whitham modulation 
theory or alternative methods, which will be discussed in detail below. 

\begin{figure}[!ht]
    \centering
    \includegraphics[width=0.40\textwidth]{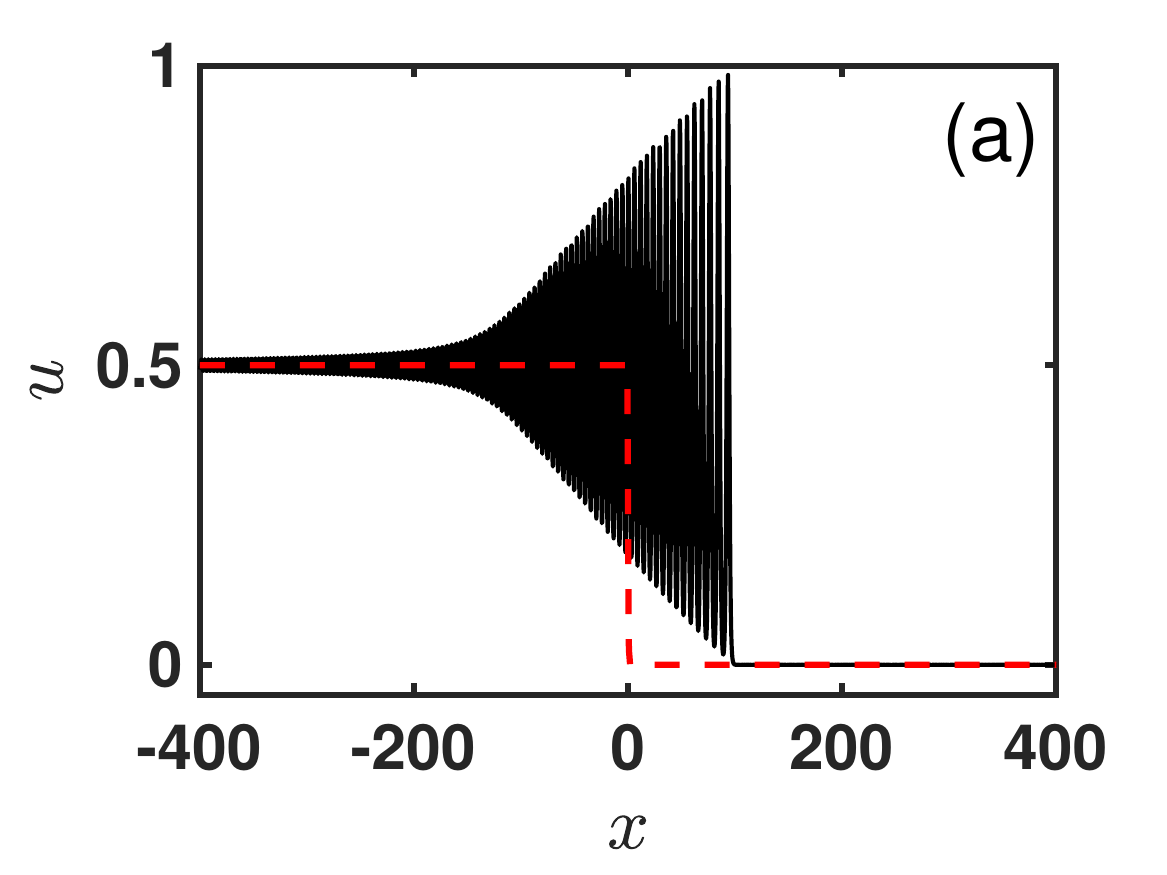}
    \includegraphics[width=0.40\textwidth]{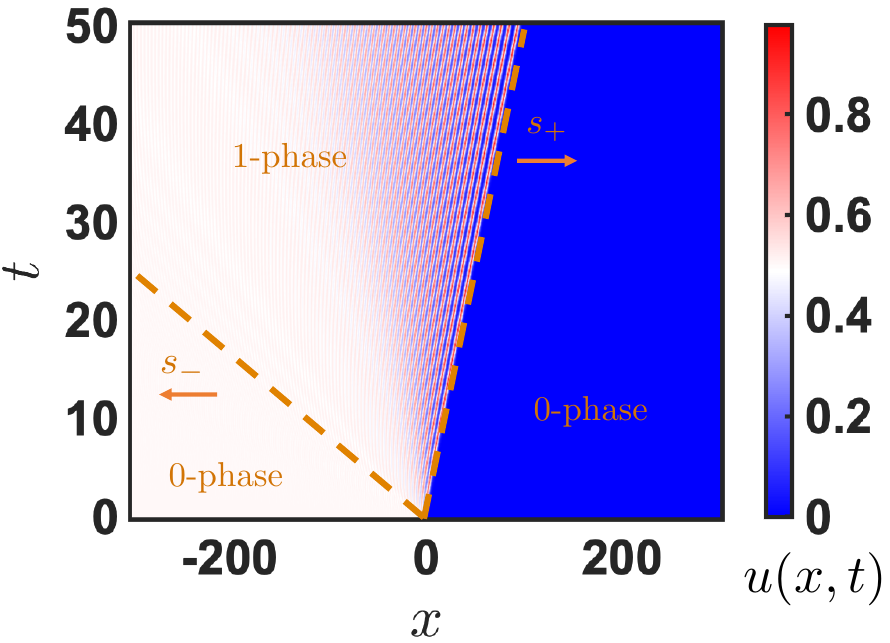}
     \includegraphics[width=0.40\textwidth]{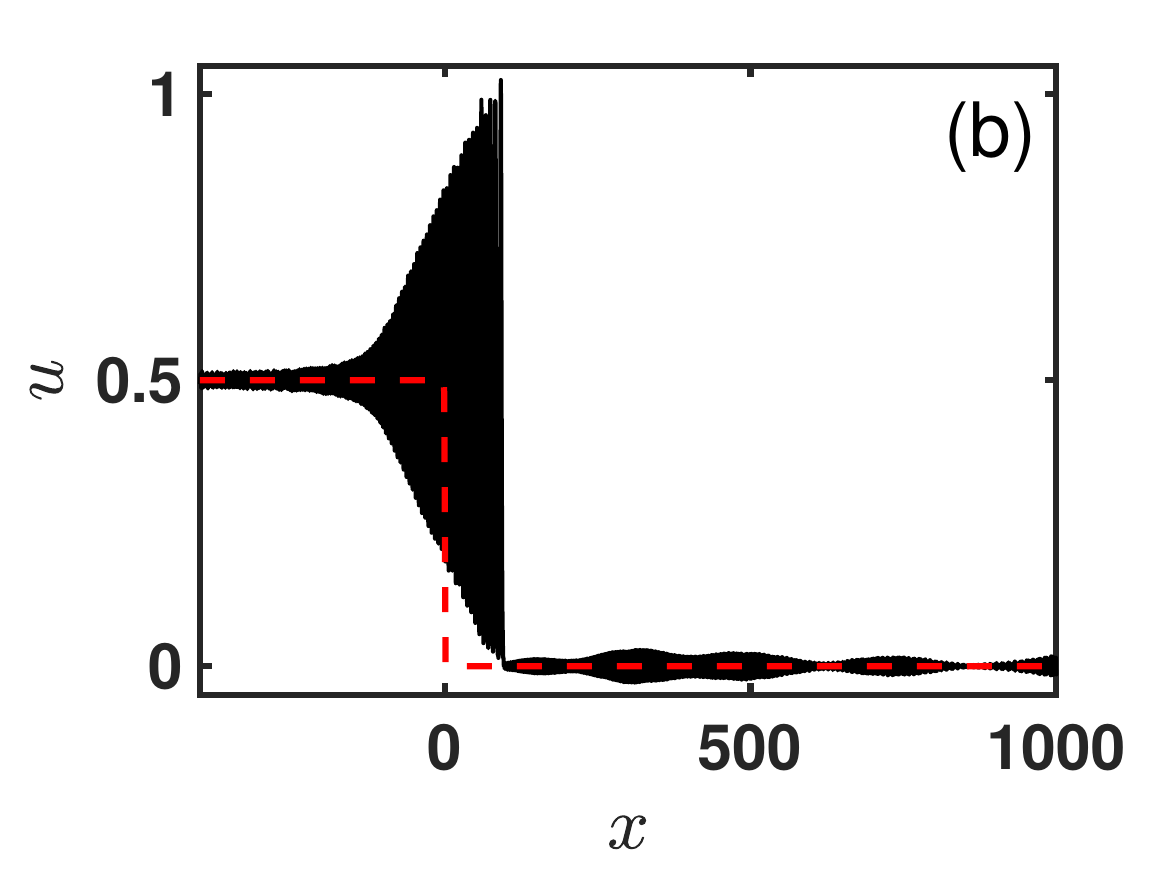}\includegraphics[width=0.40\textwidth]{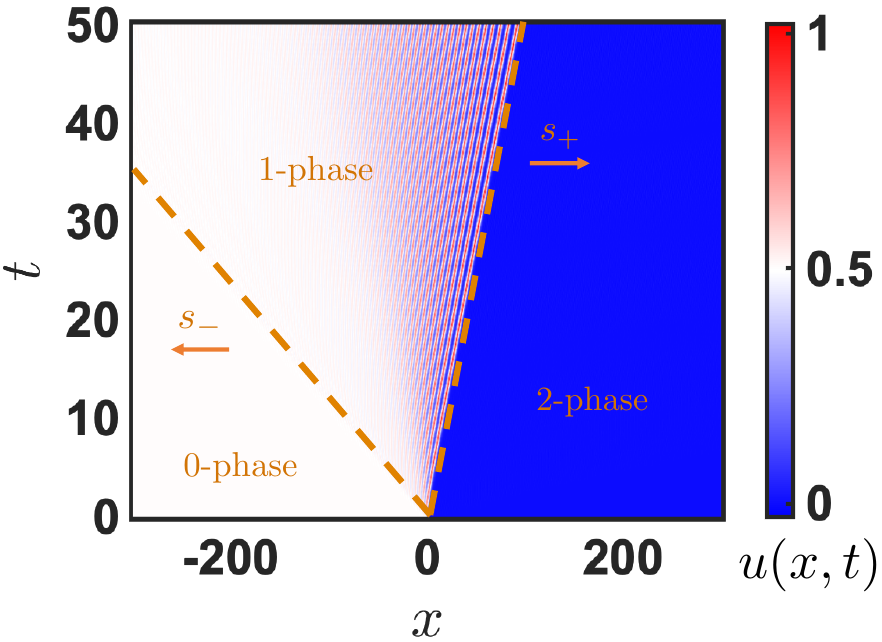}\\ \includegraphics[width=0.40\textwidth]{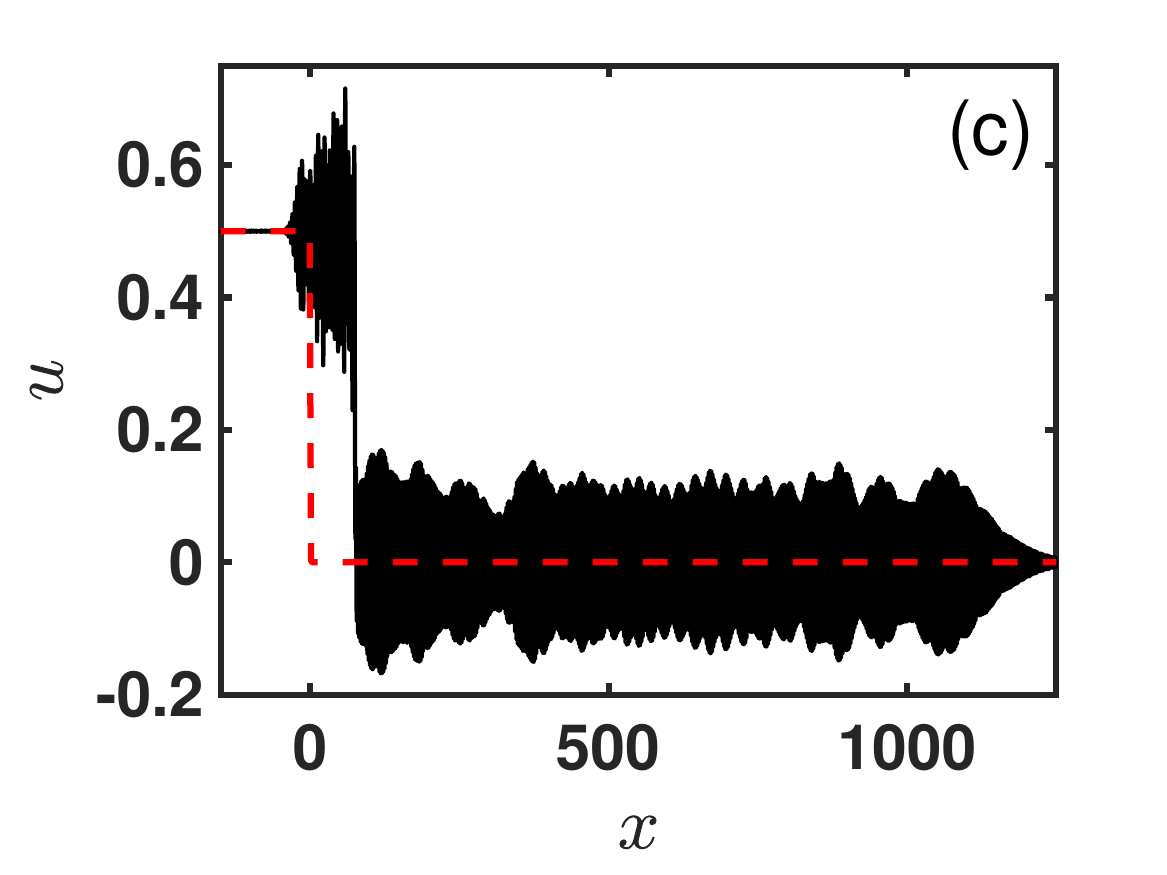}\includegraphics[width=0.40\textwidth]{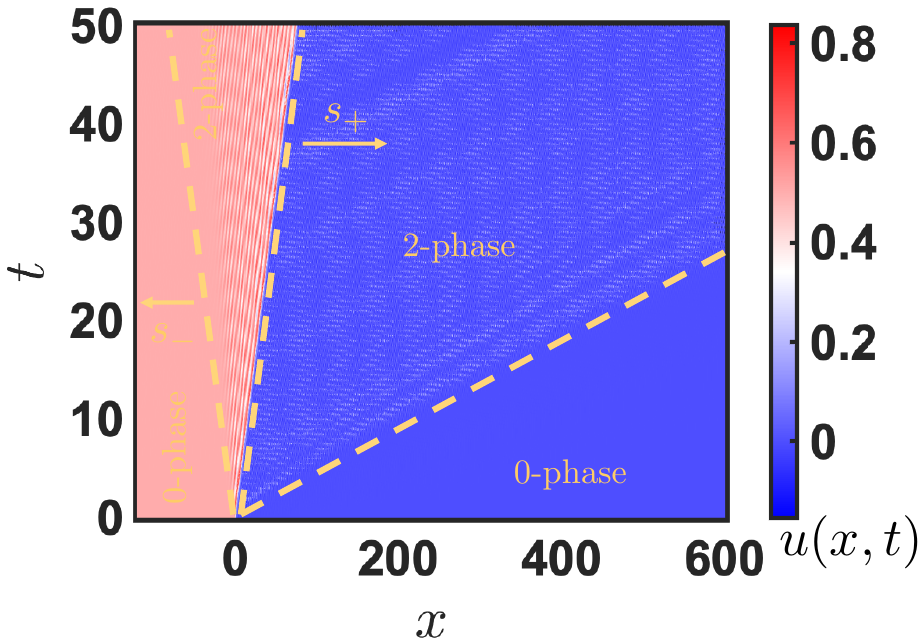}\\\includegraphics[width=0.40\textwidth]{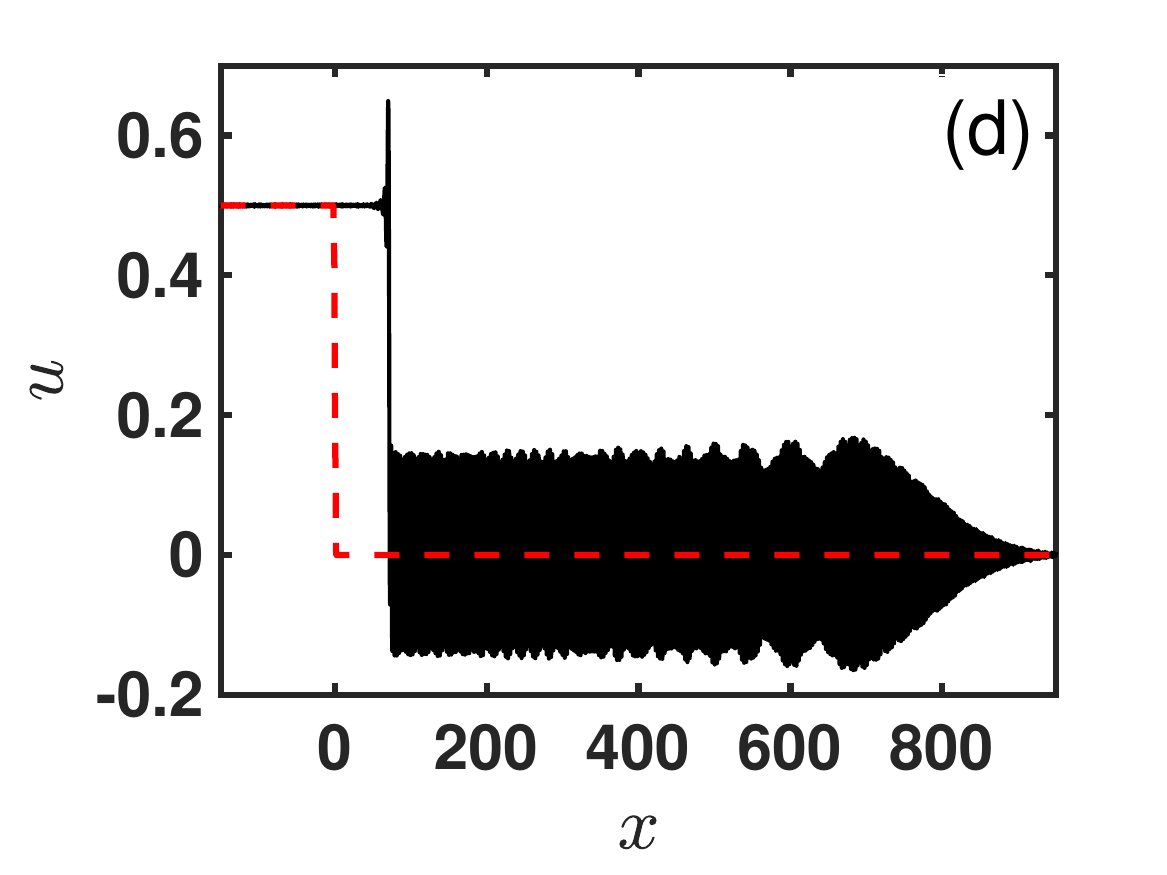}\includegraphics[width=0.40\textwidth]{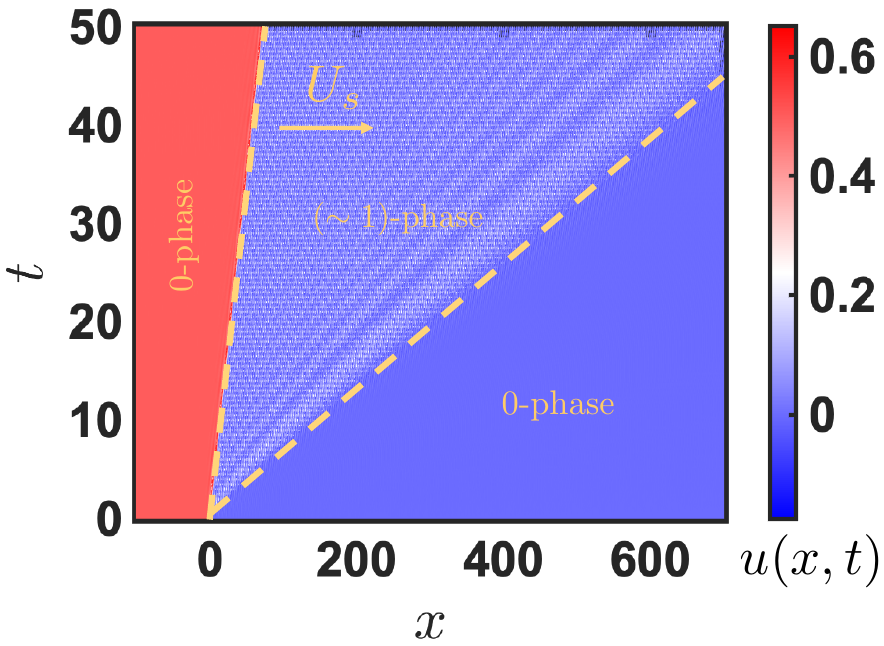}  \caption{Distinct DSW regimes with associated space-time contour plot of the solutions $u(x,t)$ in non-convex systems. Red (dashed) line marks the initial jump (\ref{ic_shock}). (a) classical KdV DSW with $c_{i}=0$, $i=1,\ldots, 4$, (b) non-classical RDSW with $c_{1}=-1$, $c_{2}=c_{3}=1$, $c_{4}=0.3$, (c) non-classical CDSW with $c_{1}=-1$, $c_{2}=c_{3}=1$, $c_{4}=1.0$, (d) non-classical TDSW with $c_{1}=-1$, $c_{2}=c_{3}=1$, $c_{4}=2.0$. Here, $t=50$, $\epsilon = 0.15$ with $\Delta=0.5$. (Color version online).}
     \label{f:dswsregimes}
\end{figure}

The formation structure of DSWs, governed by the KdV equation~(\ref{e:kdv}) or, 
more generally, the extended KdV equation~(\ref{e:ekdv}), is commonly understood to 
exist in the form depicted in Fig.~\ref{f:dswsregimes}(a). However, the arrangement 
of solitary and harmonic waves in the DSW envelope can vary based on the signs of 
the coefficient terms in the associated equation. To describe these variations, the 
concepts of DSW polarity $p$ and orientation $d$ are introduced \cite{elreview}. For 
instance, if the solitary waves are situated at the leading edge of the 
dispersive shock, the orientation of the DSW is considered positive, denoted by the 
value $d=1$; otherwise, its orientation is negative, expressed as $d=-1$. On the other 
hand, if the solitary waves are elevating waves on the varying mean level linking the 
initial steady states, the polarity of the DSW is considered positive, with the value 
$p=1$; otherwise, it has a negative polarity, indicated by $p=-1$; see Fig.~4 in  
Ref.~\cite{elreview} for a summary of these different cases.
 
The conventional perspective of dispersive shocks fundamentally changes in the context 
of dispersive hydrodynamics featuring non-convex dispersion. The loss of convexity, 
or concavity, typically occurs when higher-order terms, as seen in the extended KdV 
equation (\ref{e:ekdv}), are incorporated into the model. The introduction of non-convex dispersion allows dispersive radiation to interact with the leading solitary wave edge of the dispersive shock. In this scenario, the amplitude of the solitary wave edge decays, as well as its velocity, as it emits resonant radiation propagating ahead of it. The solitonic velocity then matches the speed of the phase velocity of the dispersive radiation. This resonance significantly influences the configuration of the DSW, causing a fundamental alteration in the classical structure described earlier. 
Near the zero dispersion point $k_i$ depicted in Fig.~\ref{f:dispersion}, 
i.e., the inflection point where the sign of curvature changes $\omega_{kk}=0$, 
the standard form of a KdV DSW becomes demolished, giving rise to novel dispersive 
hydrodynamic regimes.

Essentially, there are three non-classical, resonant shallow water dispersive shock 
regimes, depending critically on the 
magnitude of the initial jump  
$\Delta=u_{-}-u_{+}$ \cite{patkawahara,nemboreel,salehnem1,salehnem2}, 
or on the different values of the higher-order coefficients $\epsilon c_{j}$  
for a fixed initial jump \cite{pat,salehekdv}. These resonant DSWs are classified as 
{\it radiating dispersive shock wave} (RDSW), {\it cross-over dispersive shock 
wave} (CDSW), and {\it travelling dispersive shock wave} (TDSW). The descriptions of these three DSW types are as follows. 

\begin{itemize}
\item RDSW Regime: As depicted in Fig.~\ref{f:dswsregimes}(b), the DSW in this 
particular regime is similar to the standard form of a KdV DSW. However, it possesses 
distinctive features attributable to the presence of small-amplitude resonant 
radiation that is attached to the leading solitary wave edge of the bore. Each 
individual wave within the DSW additionally undergoes resonance, due to the non-
convexity effect. This resonant interaction hinders the DSW's ability to maintain a 
structured hierarchical form. The resonant radiation in the RDSW regime has a mild 
damping effect on the bore stability, which becomes more pronounced in the subsequent 
regime. 
     
\item CDSW Regime: In this regime, the bore loses its structured hierarchical form and 
experiences high modulational instability. The resonant radiation that propagates 
ahead of the bore transforms into a dispersive wave with a larger amplitude and high 
modulation, as depicted in Fig.~\ref{f:dswsregimes}(c). The damping effect imposed by 
the resonance on the bore stability is substantial within this regime. 

\item TDSW Regime: This regime is depicted in Fig.~\ref{f:dswsregimes}(d). As 
observed, the traditional structure of a DSW is completely dismantled, and the 
resonant radiation becomes a wave of significantly larger amplitude. Furthermore, the 
resonant wavetrain now maintains stability with almost uniform amplitude. This 
stability arises from the elevation of its resonant mean level, influenced by the 
varying mean level of the partial DSW propagating in front of the resonant radiation 
and taking it down to the initial level ahead $u_{+}$. In this regime, the bore is 
replaced by a partial solitary wave with negative polarity, connecting the initial 
level behind $u_{-}$ to the initial level ahead $u_{+}$. In dispersive hydrodynamics, 
this partial solitary wave structure is referred to as the Whitham shock \cite{patjump}, as we will discuss later.
    
\end{itemize}

Notice that similar non-convex DSW regimes have also been observed in the context 
of nonlinear optics of nematic liquid crystals. However, as these are beyond the 
scope of this work which deals with water waves, they will not be discussed here.  Interested readers can refer to Refs.~\cite{salehnem1,salehnem2} for more information. 

\section{Mathematical tools to analyze non-convex DSWs}

To better understand, from a theoretical point of view, the non-classical, non-convex 
dispersive hydrodynamic regimes discussed above, it is necessary to employ a variety 
of relevant methodologies. These include the Whitham modulation theory, the DSW 
fitting method, the DSW admissibility conditions, the DSW equal amplitude 
approximation, as well as the recently developed concept of Whitham shocks. These 
techniques will be elaborated upon and implemented to the eKdV model in the following 
subsections.

\subsection{Modulation theory}

In principle, analytical solutions for dispersive shocks can be obtained using modulation theory, which was developed by Whitham~\cite{whithambio2}. This method can be considered as the nonlinear counterpart to the WKB method used for linear wave oscillations \cite{holmes}, and it also serves as a partial differential equations generalization of the Krylov–Bogoliubov method applied to ordinary differential equations (ODEs) \cite{asymodes}. Modulation theory's primary objective is to derive the equations that govern the evolution of slow modulations in the wave parameters of a periodic wave solution. Whitham modulation theory can be approached in two distinct, however, related, ways: one involves the averaging of conservation laws \cite{whithampert}, while the other entails the averaging of Lagrangians \cite{whithamvar1,whithamvar2}. 
Here we will provide a short overview of both of these approaches; for a more comprehensive review and in-depth analysis, 
see Refs.~\cite{whitham,kamchatnovbook,elreview,bridgesbook,danrevisit}.

First, let us discuss the approach of averaged conservation laws.
We begin by considering an evolution equation of the general form:  
\begin{equation}
\Lambda(u,u_{x},u_{t},u_{xx},u_{tt},u_{xt},\ldots)=0.\label{e:evolution_gen}
\end{equation}
The fundamental concept in Whitham modulation theory hinges upon the existence of a periodic wave solution. To that end, we assume the presence of a $2\pi$-periodic traveling wave solution:
\begin{equation}
    u(\theta)=u(kx-\omega t).
\label{e:general_sol}
\end{equation}
Furthermore, we assume that the evolution equation (\ref{e:evolution_gen}) can be simplified to the form
\begin{equation}
u^{2}_{\theta}=F(u;\alpha_1,\alpha_2,\ldots,\alpha_{j},\ldots,\alpha_{n})=F(u;\underline{\alpha}), \quad j=1,2,\ldots,n, 
\label{e:integ}
\end{equation}
%
which is always possible for integrable equations. Here, the set of parameters 
$\left\{\underline{\alpha}\right\}$ arises as integration constants during the 
reduction of (\ref{e:evolution_gen}) to (\ref{e:integ}). These constants are 
physically linked to the wave parameters of the underlying periodic wave solution 
(\ref{e:general_sol}), such as the wavenumber, the frequency, etc. The traveling wave 
solution can be obtained by directly integrating the first-order differential equation 
(\ref{e:integ}). In nonlinear dispersive problems, this integration typically leads to 
the appearance of elliptic function solutions, such as cnoidal waves \cite{handbook}. 
Given the periodicity of the solution $u$ and the positivity of the function 
$F(u,\underline{\alpha})$, it is expected that the wave solution $u$ oscillates 
between two zeros, denoted as $u_{1}(\theta_1,\underline{\alpha})$ and 
$u_{2}(\theta_2,\underline{\alpha})$, where $u_{1}<u<u_{2}$, satisfying the equation 
$u^2_\theta=F=0$. Consequently, the corresponding wavelength (spatial period), 
wavenumber, frequency, and temporal period, are respectively given by: 
\begin{equation}
L(\underline{\alpha})=2\int^{\theta_{2}}_{\theta_{1}}d\theta=2\int^{\theta_{2}}_{\theta_{1}}
\frac{du}{u_{\theta}}=2\int^{\theta_{2}}_{\theta_{1}}\frac{du}{F(u;\underline{\alpha})},
\end{equation}
\begin{equation}
    k(\underline{\alpha}) = \frac{2\pi}{L(\underline{\alpha})},\quad \omega(\underline{\alpha}) = Vk(\underline{\alpha}), \quad T(\underline{\alpha}) = \frac{2\pi}{\omega(\underline{\alpha})},
\end{equation}
where $V$ is the phase velocity of the travelling wave solution.

Now, let us introduce the concept of slow spatial and temporal modulations in the wave parameters, which is the fundamental motivation behind the study of Whitham modulation theory. Essentially, the assumption of slow modulations implies that the wave parameters undergo minimal changes within one spatial and temporal period. Mathematically, this implies that the spatial and temporal derivatives of the wave parameters are relatively small, specifically $\partial_{x}\alpha_{j}\ll{\alpha_{j}/L}$ and $\partial_{t}\alpha_{j}\ll{\alpha_{j}/T}$. Following Whitham's approach, in order to identify the equations governing the spatio-temporal slow modulations in the wave parameters, we need to average at least $n-1$ conservation laws over the fast oscillatory scale. This can be expressed as
\begin{equation}
\frac{\partial}{\partial t}\overline{\mathcal{P}_{i}}(\underline{\alpha})+\frac{\partial}{\partial x}\overline{\mathcal{Q}_{i}}(\underline{\alpha})=0;\quad i=1,2,\ldots,n-1. \label{e:mod_system}
\end{equation}
Here, the averaged densities $\overline{\mathcal{P}}$ and averaged fluxes $\overline{\mathcal{Q}}$ implicitly depend on space $x$ and time $t$ rather than being explicitly defined. The averaging rule applied in this context is given by
\begin{equation}
\overline{\mathcal{G}}=\frac{1}{2\pi}\int^{2\pi}_{0} \mathcal{G}(\theta;u,\underline{\alpha})d\theta.
\end{equation}
To complete the Whitham modulation system (\ref{e:mod_system}) with an $n$-th equation, we consistently employ the modulation relation, in larger scale, between the wavenumber $k=\theta_{x}$ and frequency $\omega=-\theta_{t}$ by 
using the compatibility condition $\theta_{xt}=\theta_{tx}$; this yields
the so-called {\it conservation of waves equation}, 
or the {\it consistency equation}:
\begin{equation}
k_{t}+\omega_{x}=0.
\label{e:closure}
\end{equation}

Second, we discuss the variational approach based on averaged Lagrangians to modulation theory. With some Lagrangian $L(u,u_{x},u_{t},u_{xx},u_{tt},u_{xt},\ldots)$ that does not depend explicitly on space and time, let us assume that the evolution equation (\ref{e:evolution_gen}) satisfies the Euler-Lagrange equation 
\begin{equation}
    L_{u}- \frac{\partial }{\partial t}L_{u_t} - \frac{\partial }{\partial x}L_{u_x} + \frac{\partial^2 }{\partial t^2}L_{u_tt} +  \frac{\partial^2 }{\partial x^2}L_{u_{xx}} + \cdots = 0,
\end{equation}
which results from the principle of least action (Hamilton's principle):
\begin{equation}
    \delta \int\int L(u,u_{x},u_{t},u_{xx},u_{tt},u_{xt},\ldots) dxdt=0.
\end{equation}
Whitham proposed an alternative method to study wave modulations as follows. First, and as previously stressed, the main ingredient in modulation theory is the existence of a periodic wave solution. Typically, in water wave theory, or nonlinear problems in general, the corresponding Lagrangian appears only in terms of the derivatives of a potential function \cite{whitham,kamchatnovbook}, say, $u=\phi_{x}$. Thus, the Lagrangian reads $L(\phi_{x},\phi_{xx},\phi_{xt},\ldots)$, and we consider the following generic form of a uniform wavetrain \cite{whitham,whithamvar2}:
\begin{equation}
    \phi=\bar{u}x-\gamma t +\Phi(\theta,a), \label{e:gen_uniform}
\end{equation}
where $\bar{u}$ is the mean flow variable, $\gamma$ is some constant (termed pseudo-frequency), $\theta=kx-\omega t$ is the uniform phase, and $a$ denotes the amplitude of the wave. The function $\Phi$ represents a wavetrain propagating on a zero mean level. Now, when slow modulations in space and time are taken into account, 
Eq.~(\ref{e:gen_uniform}) can be generalized by 
introducing the pseudo-phase variable $\psi$ \cite{whitham,whithamvar2}, such that:
\begin{equation}
    \phi=\psi + \Phi(\theta,a),
\end{equation}
and 
\begin{equation}
    \psi_{x}=\bar{u},\quad \psi_{z}=-\gamma,\quad \theta_{x}=k,\quad \theta_{t}=-\omega.
\end{equation}
In water wave theory, the variables $\bar{u}$ and $\gamma$ play important roles in representing the mean level and the mean fluid velocity. In the context of modulations, the function $\psi$ gives the mean level variation for the wavetrain governed by the function $\Phi$. 

Whitham argued that the slow evolution is governed by the (averaged) Hamilton's principle \cite{whithamvar1,whithamvar2}:
\begin{equation}
     \delta \int\int \mathcal{L}(\psi_{x},\psi_{t},\theta_{x},\theta_{t},a) dxdt=0,
\end{equation}
where the averaged Lagrangian $\mathcal{L}$ is defined by 
\begin{equation}
 \mathcal{L}(\psi_{x},\psi_{t},\theta_{x},\theta_{t},a) =\frac{1}{2\pi}\int^{2\pi}_{0} \mathcal{L}(\theta; \psi_{x},\psi_{t},\theta_{x},\theta_{t},a) d\theta.  
\end{equation}
Then, it follows that the set of equations that govern the slowly varying wave parameters are the Euler-Lagrange equation deduced by taking variations of $\mathcal{L}$ with respect to the wave parameters. The variations with respect to $a,$ $\theta,$ and $\psi$ give, respectively, 
\begin{eqnarray}
& &  \mathcal{L}_{a} = 0, 
\label{e:gen_mod_eq1}\\
& &  \frac{\partial}{\partial t}\mathcal{L}_{w}-\frac{\partial}{\partial x}\mathcal{L}_{k} =  0, 
\label{e:gen_mod_eq2}\\
& &  \frac{\partial}{\partial t}\mathcal{L}_{\gamma}-\frac{\partial}{\partial x}\mathcal{L}_{\bar{u}} = 0.
\label{e:gen_mod_eq}
\end{eqnarray}
The first averaged Euler-Lagrange equation gives the associated nonlinear dispersion relation $\omega=\omega(k,a)$, the second one leads to conservation of wave action (analogous to an adiabatic invariant of classical mechanics), and the third equation determines the mean flow modulation, and typically yields the conservation of mass. As previously explained, the closure of the modulation system is always done by taking the consistency equation (\ref{e:closure}). 

Note that when the periodic traveling wave solution is simply $u=u(\theta,a)$ and there is no appearance of a potential function $\phi$ in the Lagrangian $L$, then the associated averaged Lagrangian is $\mathcal{L}(\omega,k,a)$ and the corresponding modulation equations are 
(\ref{e:gen_mod_eq1}) and (\ref{e:gen_mod_eq2}). Therefore, the above variational formulation is an extension of modulation theory to more variables.

Conservation laws in terms of the averaged Lagrangian $\mathcal{L}$ that result 
from N\"{o}ether's theorem \cite{fomin} also play crucial roles in various non-convex dispersive hydrodynamics problems ---see, e.g., Refs.~\cite{pat,salehnem1} for water waves and nonlinear optics applications. 
The form of the Lagrangian $\mathcal{L}$ permits the derivation of the energy 
conservation and the momentum conservation, due to the explicit independence of $x$ 
and $t$. Exploiting the symmetry of $\mathcal{L}$ with respect to time translation 
$t^*=t+t_{0}$, that is invariance in time, yields 
\begin{equation}
    \frac{\partial}{\partial t}\left(\omega \mathcal{L}_{\omega} + \gamma \mathcal{L}_{\gamma} - \mathcal{L}\right) - \frac{\partial}{\partial x}\left(\omega \mathcal{L}_{k} - \gamma \mathcal{L}_{\bar{u}} \right)=0,
\end{equation}
which is the equation for the energy conservation. Similarly, the invariance 
of $\mathcal{L}$ in space, $x^*=x+x_{0}$, leads to: 
\begin{equation}
    \frac{\partial}{\partial t}\left(k\mathcal{L}_{\omega} + \bar{u} \mathcal{L}_{\gamma}\right) - \frac{\partial}{\partial x}\left(k \mathcal{L}_{\bar{u}} + \bar{u} \mathcal{L}_{\bar{u}} - \mathcal{L} \right)=0,
\end{equation}
which is the equation for momentum conservation. The viewpoint of averaged 
Lagrangians ---and also averaged conservation laws--- to modulation theory can be fully justified using the method of multiple scales from perturbation theory \cite{whitham,lukemulti,ablowitzmulti}. The modulation equations are found to be exactly the conditions needed to eliminate secular terms. 

Having determined the modulation equations, it is possible to investigate solutions for dispersive shocks. To do this, the modulation equations need to be set 
in a system of first-order quasi-linear PDEs of the form:
\begin{equation}
 \mathbf{v}_{t} + \mathcal{B}(\mathbf{v}) \mathbf{v}_{x} = \mathbf{0},\label{e:quasi_sys}
\end{equation}
where $\mathbf{v}^{T}=[v_1\,v_2\ldots v_n]$ is a differentiable vector connected to 
the slowly varying wave parameters and $\mathcal{B}$ is a non-singular matrix. In 
the cases of averaged conservation laws and averaged Lagrangians methods discussed above, we have $\mathbf{v}^{T}=[\alpha_{1}\,\alpha_{2}\ldots\alpha_{n}]$ 
and $\mathbf{v}^{T}=[\bar{u}\,a\,k\,\gamma]$, respectively. If the associated eigenvalues of this system are real (real and distinct), then the Whitham modulation system is hyperbolic (strictly hyperbolic) and the underlying periodic wave solution is modulationally stable. On the other hand, if the associated eigenvalues are purely imaginary, then the Whitham modulation system is elliptic and the underlying periodic wavetrain is modulationally unstable. 

The breakthrough after the development of modulation theory was the seminal work 
of Gurevich and Pitaevskii \cite{plasmapit} where it was realized 
that if the quasi-linear system~(\ref{e:quasi_sys}) is hyperbolic and can be set in Riemann invariant form, namely:
\begin{equation}
   \frac{\partial r_{j}}{\partial t} + \lambda_{j}(r_{1},r_2,\ldots,r_{n}) \frac{\partial r_{j}}{\partial x}=0\quad\text{on the characteristics}\quad\frac{d x}{d t}=\lambda_{j},\label{e:RI}
\end{equation}
where $r_{j}$ ($j=1,2,\ldots,n$) are the constants along the characteristics 
(so-called {\it Riemann invariants}) and $\lambda_{j}$ are the associated eigenvalues (representing group velocities), then the set of characteristics that form a simple wave solution correspond to a DSW solution (see also 
Ref.~\cite{elreview} for DSW solutions arising in   
a wide class of hydrodynamic problems). For the KdV equation (\ref{e:kdv}), the set of characteristics that give a simple fan solution, that is, a DSW, are 
\begin{equation}
    \frac{x}{t}=6u_{+}+2\Delta(1+m)-4\Delta\frac{m(1-m)K(m)}{E(m)-(1-m)K(m)},
\end{equation}
where $K(m)$ and $E(m)$ are the elliptic integrals of first and second kind, and
$0\leq{m}\leq{1}$ is the elliptic modulus \cite{handbook}. The leading solitary wave 
edge corresponds to $m\to{1}$, whereas the trailing harmonic wave edge 
corresponds to $m\to{0}$. Therefore, the leading solitary wave edge propagates 
with velocity
\begin{equation}
    s_{+}=6u_{+}+4\Delta, 
    \label{e:kdvsolvel}
\end{equation}
and the trailing harmonic edge propagates with the group velocity 
\begin{equation}
    s_{-}=6u_{-}-12\Delta. 
    \label{e:kdvtrail}
\end{equation}

The determination of Riemann invariants, however, is always a guarantee for $2\times2$ 
systems \cite{whitham} or for nonlinear dispersive systems which are integrable via 
the method of the inverse scattering transform or finite-gap integration theory 
\cite{kamchatnovbook,analysis}. In practice, most of nonlinear dispersive wave 
equations that arise in applications are non-integrable. It is then a challenge to 
find dispersive shock solutions for such type of problems, as the existence of an 
exact periodic wave solution \eqref{e:general_sol} and a Riemann invariant form 
(\ref{e:RI}) are impossible. This urges the necessity to develop alternative, 
approximate methods to find DSW solutions governed by non-integrable problems, such as 
the eKdV equation (\ref{e:ekdv}) or the Kawahara equation (\ref{e:kawahara}). 

Among these methods are the DSW fitting method, which is based solely on the
knowledge of corresponding linear dispersion relation, and the DSW equal amplitude 
approximation, which is based on the knowledge of corresponding conservation laws and 
a solitary wave solution. These methods will be detailed below, in the Subsections 
\ref{subsec:gennady} and \ref{subsec:equal}. For the case when a non-integrable 
equation is composed of a classical integrable equation plus small asymptotic 
corrections, one can approach the problem by employing a transformation that maps the 
non-integrable equation to the integrable one. This is key to the concept of 
asymptotic integrability \cite{fokas,kodama} and, importantly, also allows for the approximation of  
a DSW solution of a non-integrable model 
by a perturbed DSW solution of the corresponding integrable one.  
A pertinent example, relevant to our study, is the eKdV equation (\ref{e:ekdv}) 
and its special integrable case (\ref{e:kdv}). Indeed, in this case, the nonlocal transformation \cite{ekdhighermodu}:
\begin{equation}
  u=\eta + \epsilon \left[c_{5}\eta^{2} + c_{6}\eta_{xx} + c_{7}\eta_{x}\int^{x}_{v_{p}t}\left(\eta(x',t)-\bar{u}\right)dx' \right]
\end{equation}
maps (\ref{e:ekdv}) to the KdV equation in the standard form 
\begin{equation}
\eta_{\tau}+6\eta\eta_{\xi}+\eta_{\xi\xi\xi}=0,
\label{e:kdv_scaled}
\end{equation}
with $\tau= t+(\epsilon c_4/3)x$ and 
$\xi=x+\epsilon\left[c_{7}\bar{u}(x-v_{p}t)+c_{7}\gamma t\right]$, 
where $\gamma$ is the first integration constant to the KdV 
Eq.~(\ref{e:kdv_scaled}). Here, $v_{p}$ and $\bar{u}$ are the phase velocity and the background of the KdV periodic cnoidal wave solution, respectively. 
Moreover, $c_{5}=(c_3-c_1+4c_4)/6$, $c_6=(c_2-6c_4-c_1)/12$ and 
$c_{7}=(8c_{4}-c_{3})/3$. The associated modulated mean level, amplitude, wavenumber, leading solitary wave velocity, and trailing harmonic edge velocity, are 
respectively given by:
\begin{eqnarray}
    \bar{u}& = & u_{+}-\Delta + \Delta\left(2\frac{E(m)}{K(m)}+m\right)+\epsilon \frac{\Delta^2}{3}\left[c_{5} \left(2-5m+3m^2+(4m-2)\frac{E(m)}{K(m)}\right) \right. 
    \nonumber \\
    & & \left. \mbox{} + 4c_{7}\left(3\left(1-\frac{E(m)}{K(m)}\right)^2 - (2+2m)\left(1-\frac{E(m)}{K(m)}\right)+m \right) \right], \label{e:extended_mean} \\
    a & = & 2m\Delta +\epsilon\Delta^2\left[c_5 m + 2c_6 (m^2-2m) \right], \\
    k & = & \frac{\pi\Delta^2}{K(m)}\left[1+\epsilon\left(c_7 \left(4u_{-}u_{+}-\Delta^2m^2 + 2\Delta u_{-}m-u^2_{-}\right) -\frac{1}{3}c_{4}\left( 2\Delta m + 2u_{-}+4u_{+} \right) \right. \right. 
    \nonumber \\
    & & \left. \left. \mbox{} - \frac{1}{2}c_{5}\left(\Delta+2u_{+}\right)
       \right) \right], \\
    s_{+} & = & 6u_{+}+4\Delta + \epsilon\left[ u^{2}_{-}\left(\frac{16}{3}c_{4}-4c_{5}\right) + u_{-}u_{+}\left(\frac{16}{3}c_{4}-4c_{7}\right)  \right. 
    \nonumber \\
    & & \left. + u^{2}_{+}\left(\frac{4}{3}c_{4}-2c_{5}+c_{7}\right) \right], \label{e:extended_leading} \\
    s_{-} & = & 6u_{-}-12\Delta + \epsilon \left[u^{2}_{-}\left(12c_{4}+6c_{5}+9c_{7}\right) - u_{-}u_{+}\left(12c_{7}+\frac{144}{3}c_{4}\right)  \right. 
    \nonumber \\
    & & \left. + u^{2}_{+}\left(\frac{144}{3}c_{4}-12c_{5}\right) \right].\label{e:extended_trailing} 
\end{eqnarray}
Notice that, corrections have been made to address several typographical errors 
in the asymptotic expressions presented in Ref.~\cite{ekdhighermodu}, 
which led, accordingly, to 
Eqs.~(\ref{e:extended_mean})–(\ref{e:extended_trailing}). 

In the context of eKdV non-convex dispersive hydrodynamics, the higher order 
modulation theory wave parameters (\ref{e:extended_mean})–(\ref{e:extended_trailing}) 
holds valid and are effective in the case of the RDSW regime ---see 
Fig.~\ref{f:dswsregimes}(b)--- as the resonant radiation is relatively small. 
However, they cease to be valid as the resonant radiation amplitude becomes 
large as in the CDSW and TDSW regimes.

\subsection{Dispersive shock fitting method}
\label{subsec:gennady}

In the absence of integrability (thus, lacking an inverse scattering transform 
solution) and Whitham modulation equations (hence, lacking a Riemann invariant form), 
the dispersive shock fitting method, pioneered by El \cite{fitting}, proves effective 
for stable KdV-type DSWs. This method enables the determination of the macroscopic 
properties of the DSW solely through the associated \textit{linear} dispersion 
relation $\omega(k;\bar{u})$. It provides predictions for both the 
velocity of the leading solitary wave edge and the velocity of the trailing harmonic 
edge, as well as the leading solitary wave amplitude in the case of the presence of a 
velocity-amplitude relation. The foundation of this method, achieved without requiring 
detailed knowledge of the full modulation equations, stems from the observation of the 
coalescence of the modulation equations at the leading solitary wave edge and the trailing harmonic edge. It is worth mentioning that the fitting method has recently been extended 
in Ref.~\cite{fittinginterior}, using asymptotic analysis based on nonlinear 
Schr\"{o}dinger (NLS) equation and its higher order approximation, to capture the 
microstructure of a DSW, and specifically its interior. 
Extending the application of this method to the eKdV equation 
is beyond the scope of this work, and will not be addressed here, 
as it is a subject of future study. 

To explain the essence of this method,  
we consider the classical KdV equation (\ref{e:kdv}), whose linear dispersion relation 
is $\omega(k;\bar{u})=6\bar{u}k-k^3$. At the solitary wave edge the wavenumber becomes 
$k=0$, so that there remain only two variables, the mean level $\bar{u}$ and the 
solitary wave amplitude $a$. Then, for consistency, two of the Riemann invariants 
and their associated characteristics become coalescing. At the trailing, linear 
wave edge, the amplitude $a$ vanishes, so that there remain only two variables, the 
mean level $\bar{u}$ and the wavenumber $k$. Again, for consistency, two Riemann 
invariants and their associated characteristics become coalescing. 

Let us begin by examining the trailing harmonic wave edge. Behind the DSW, the solution behaves as non-dispersive. Consequently, the KdV equation (\ref{e:kdv}) simplifies to the Hopf equation for the mean level $\bar{u}$,
\begin{equation}
 \bar{u}_{t} + 6 \bar{u}\bar{u}_{x} = 0.
 \label{e:meankdv}
\end{equation}
In the linear limit, the mean level decouples from the amplitude. At the trailing harmonic edge, the solution is entirely determined by the mean level there, with $k = k(\bar{u})$, as $a=0$ at the trailing edge and remains fixed. Thus, we have 
\begin{equation}
 \frac{d k}{d \bar{u}} \frac{\partial \bar{u}}{\partial t} + \frac{\partial \omega}{\partial \bar{u}} \frac{\partial \bar{u}}{\partial x} + \frac{\partial \omega}{\partial k} \frac{dk}{d\bar{u}} \frac{\partial \bar{u}}{\partial x} = 0,
 \label{e:compa1}
\end{equation}
which can be rewritten in a more convenient form as 
\begin{equation}
 \bar{u}_{t} + \left[ \frac{\omega_{\bar{u}} + \omega_{k}\, k'(\bar{u})}{k'(\bar{u})} \right] \bar{u}_{x} = 0. 
 \label{e:comp2}
\end{equation}
For the mean level equations (\ref{e:meankdv}) and (\ref{e:comp2}) to be compatible, the ratio between the square brackets in (\ref{e:comp2}) must be equal to $6\bar{u}$. This leads to the ODE
\begin{equation}
     \frac{d k}{d\bar{u}} = \frac{\omega_{\bar{u}}}{6\bar{u}-\omega_{k}} = \frac{2}{k}, \label{e:trailode}
\end{equation}
which, using the KdV linear dispersion relation, yields integral curve solutions satisfying 
\begin{eqnarray}
 k^{2} =  4\bar{u} + k_{o}.
\end{eqnarray}
To determine the constant $k_{o}$, a boundary condition at the leading solitary wave edge of the DSW is necessary. At the leading edge, the mean level is $\bar{u} = u_{+}$ and the wavenumber is $k=0$. Hence, the wavenumber in terms of the mean level is
\begin{equation}
 k^{2}(\bar{u}) = 4\left( \bar{u} - u_{+} \right),
 \label{e:ksoln}
\end{equation}
resulting in the harmonic wavenumber $k(\bar{u})=k(u_{-})=k_{-}$ satisfying 
\begin{equation}
 k_{-}^{2} = 4 \left( u_{-} - u_{+} \right)=4\Delta.
 \label{e:kminus}
\end{equation}
Thus, the trailing harmonic edge of the DSW propagates at the group velocity 
\begin{equation}
 s_{-} = 6 u_{-} - 3 k_{-}^{2} = 6u_{-}-12\Delta,
 \label{e:cglin}
\end{equation}
which is consistent with (\ref{e:kdvtrail}).

The analysis at the leading solitary wave edge is more intricate and necessitates an 
understanding of some properties of elliptic functions. It hinges on the observation 
that the KdV cnoidal wave solution is represented by the elliptic function $\cn^{2}$, 
which exhibits double periodicity with period $2K(m)$ in the real direction and 
$2K'(m)$ in the imaginary direction. Consequently, the solitary wave demonstrates 
periodicity in the imaginary direction. Furthermore, the dispersion relation resembles 
the linear dispersion relation in the imaginary direction. Specifically, the KdV 
solitary wave solution of amplitude $a_{+}$ traveling on the mean level $\bar{u}$ is 
given by \cite{whitham,kamchatnovbook}
\begin{equation}
 u = a_{+}\sech^{2} w_{s}\theta_{s},
 \label{e:kdvsol}
\end{equation}
where $w_{s}=\sqrt{a_{+}/2}$ denotes the inverse width, $\theta_{s}=x-s_{+}t$ the 
phase, with the velocity given by $s_{+}=6\bar{u}+2a_{+}$. As $x \to \infty$, the 
solitary wave solution tends to
\begin{equation}
 u \to 4a e^{-\sqrt{2a_{+}} \left( x - s_{+}t \right)}.
\label{e:sollim}
\end{equation}
Upon setting $\widetilde{k} = \sqrt{2a_{+}}$, the limit solution (\ref{e:sollim}) transforms into 
\begin{eqnarray}
 u \to  4a e^{-\left( \widetilde{k}x - \widetilde{\omega}t \right)},
 \label{e:sollim2}
\end{eqnarray}
where $\widetilde{\omega} = -i\omega(i\widetilde{k};\bar{u}) =6\bar{u} 
\widetilde{k} + \widetilde{k}^{3}$. Subsequently, setting $\widetilde{k} = -ik$, yields 
\begin{equation}
 u \to 4a e^{i(kx - \omega t)},
 \label{e:sollindisp}
\end{equation}
which corresponds to a harmonic wave train. Following El \cite{fitting}, 
$\widetilde{k}$ is referred to as the conjugate wavenumber and $\widetilde{\omega}$ 
as the solitonic conjugate linear dispersion relation. Essentially, the conjugate 
wavenumber $\widetilde{k}$ serves as an amplitude-type variable.

At this point, we follow the same compatibility of the modulation equations 
as for the linear edge of the bore. This leads to (\ref{e:trailode}), wherein 
the wavenumber and frequency are replaced by their conjugate equivalents, namely:
\begin{equation}
 \frac{d\widetilde{k}}{d\bar{u}} = \frac{\widetilde{\omega}_{\bar{u}}}{6\bar{u}-\widetilde{\omega}_{\widetilde{k}}} = -\frac{2}{\widetilde{k}}.
 \label{e:gennadylinsol}
\end{equation}
Utilizing the boundary $\widetilde{k}=0$ at the trailing harmonic edge $\bar{u}=u_{-}$, the solution to the ODE (\ref{e:gennadylinsol}) satisfies 
\begin{equation}
    \widetilde{k}^2(\bar{u})=4(u_{-}-\bar{u}),
\end{equation}
resulting in the conjugate wavenumber at the leading solitary wave edge being 
\begin{equation}
    \widetilde{k}^2_{+}=4(u_{-}-u_{+})=4\Delta. 
\end{equation}
It is noteworthy that $\widetilde{k}^2_{+}$ and $k^2_{-}$ are identical in the context 
of the KdV equation; however, it is not necessarily a universal characteristic 
applicable to all dispersive hydrodynamic systems. Consequently, the velocity of the 
leading solitary wave edge of the DSW is given by:
\begin{equation}
 s_{+} = \frac{\widetilde{\omega}}{\widetilde{k}} = 6u_{+}+\widetilde{k}^2 = 6u_{+}+4\Delta,
 \label{e:velsol}
\end{equation}
which aligns with the modulation theory solution (\ref{e:kdvsolvel}). As for the 
leading solitary wave edge height of the DSW, we derive it by exploiting the 
velocity-amplitude relation $s_{+} = 6\bar{u} + 2a_{+}$ for the KdV solitary wave 
solution. At the leading solitary wave edge $\bar{u}=u_{+}$, this relation yields
\begin{equation}
    a_{+}=2\Delta,
\end{equation}
with the use of (\ref{e:velsol}), consistent with modulation theory.

To summarize, for a general dispersive hydrodynamic system with an associated linear dispersion relation $\omega(k;\bar{u})$, the properties of the trailing harmonic edge and the leading solitary wave edge of a KdV-type DSW can be analyzed by solving the boundary value problems:
\begin{equation}
\frac{d k}{d\bar{u}}=\frac{\omega_{\bar{u}}}{V(\bar{u})-\omega_{k}},\quad \text{subject to}\quad k(u_{+})=0,
\end{equation}
\begin{equation}
\frac{d \widetilde{k}}{d\bar{u}}=\frac{\widetilde{\omega}_{\bar{u}}}{V(\bar{u})-\widetilde{\omega}_{\widetilde{k}}},\quad \text{subject to}\quad \widetilde{k}(u_{-})=0,
\end{equation}
respectively. Here, $V(\bar{u})$ represents the wave velocity for the corresponding  hydrodynamic equation(s) when the dispersive terms are disregarded. At the hyperbolic systems level, this corresponds to the characteristic velocity on the positive characteristic curves $C_{+}$ associated with the Riemann invariant form of the non-dispersive hydrodynamic system. The trailing edge of the DSW propagates with the group velocity:
\begin{equation}
s_{-}=\frac{\partial \omega(k_{-};u_{-})}{\partial k},
\end{equation}
while the leading solitary wave edge propagates with the velocity:
\begin{equation}
s_{+}=\frac{\widetilde{\omega}\left(\widetilde{k}_{+};u_{+}\right)}{\widetilde{k}}.
\end{equation}

Let us extend our illustration of the fitting technique to encompass eKdV DSWs, given its central role in this paper. The equations governing the fitting of the trailing edge and lead solitary wave edge are as follows:
\begin{equation}
\frac{d k}{d\bar{u}}=\frac{6k+\epsilon\left(2c_{1}\bar{u}k-c_{3}k^3\right)}{3k^{2}+\epsilon\left(3c_{3}\bar{u}k^{2}-5c_{4}k^{4}\right)},\quad \text{subject to}\quad k(u_{+})=0,
\label{e:ekdv_fitting_trailing}
\end{equation}
\begin{equation}
\frac{d \widetilde{k}}{d\bar{u}}=-\frac{6\widetilde{k}+\epsilon\left(2c_{1}\bar{u}\widetilde{k}+c_{3}\widetilde{k}^3\right)}{3\widetilde{k}^{2}+\epsilon\left(3c_{3}\bar{u}\widetilde{k}^{2}+5c_{4}\widetilde{k}^{4}\right)},\quad \text{subject to}\quad \widetilde{k}(u_{-})=0,
\label{e:ekdv_fitting_leading}
\end{equation}
Here, the dispersionless velocity is  $V(\bar{u})=6\bar{u}+\epsilon c_{1}\bar{u}^{2}$. 
Unlike the KdV equation and the Kawahara equation (\ref{e:kawahara}), theoretical 
solutions for the eKdV fitting ODEs cannot be directly derived, to the best of our 
attempts, and the equations must be integrated numerically. Ideally, this can be done 
by implementing \textit{forward} Runge-Kutta method of 4th-order (RK4) at the trailing 
edge and \textit{backward} RK4 method at the leading solitary wave edge. However, one 
can exploit the small size of the parameter $\epsilon$ to undertake some asymptotics 
and derive approximate solutions. 

Let us begin with the analysis of the harmonic trailing edge of the DSW. By employing a Taylor expansion to 
the fraction in (\ref{e:ekdv_fitting_trailing}), we acquire
\begin{equation}
   \frac{d k}{d\bar{u}} = \frac{2}{k} + \epsilon \left[ \frac{(10c_{4}-c_{3})k^2 - 2\bar{u}(3c_{3}-c_{1})}{3k} \right] +\mathcal{O}(\epsilon^2). 
\end{equation}
This reduced nonlinear ODE can be analytically solved, and its trailing harmonic wavenumber solution satisfies
\begin{eqnarray}
k^{2}(\bar{u})&=&k_{o}\exp{\left(-\frac{2}{3}\bar{u}\epsilon\left(c_{3}-10c_{4}\right)\right)} 
\nonumber \\
&+& \frac{\left(15c_3-3c_1-60c_4\right)+\epsilon\bar{u}\left(2c_3c_1-20c_4c_1-6c^2_3+60c_4c_3\right)}{\epsilon\left(c^2_3-20c_3c_4+100c^2_4\right)}.
\end{eqnarray}
By utilizing the boundary condition at the leading solitary wave edge $k(u_{+})=0$, we can determine the constant of integration as
\begin{eqnarray}
   k_{o}&=&\exp{\left(\frac{2}{3}u_{+}\epsilon\left(c_{3}-10c_{4}\right)\right)} 
\nonumber \\   
   &-& \frac{\left(15c_3-3c_1-60c_4\right)+\epsilon u_{+}\left(2c_3c_1-20c_4c_1-6c^2_3+60c_4c_3\right)}{\epsilon\left(c^2_3-20c_3c_4+100c^2_4\right)}.
\end{eqnarray}
Consequently, the trailing edge travels with the group velocity
\begin{equation}
   s_{-}=(6u_{-}+\epsilon c_{1}u_{-}^{2})
   -3\left(1+\epsilon c_{3}u_{-}\right)k^2_{-} + 5\epsilon c_{4} k^4_{-}.
\end{equation}
Similarly, the analysis of the leading solitary wave edge follows a comparable approach. Expanding the fraction in (\ref{e:ekdv_fitting_leading}) using Taylor expansion results in
\begin{equation}
   \frac{d \widetilde{k}}{d\bar{u}} = -\frac{2}{\widetilde{k}} + \epsilon \left[ \frac{(10c_{4}-c_{3})\widetilde{k}^2+2\bar{u}(3c_{3}-c_{1})}{3\widetilde{k}} \right] +\mathcal{O}(\epsilon^2). 
\end{equation}
Again, and as for the trailing edge, this ODE can be integrated analytically, yielding the conjugate wavenumber at the solitary wave edge of the DSW, which satisfies: 
\begin{eqnarray}
    \widetilde{k}^{2}(\bar{u})&=&k_{o}\exp{\left(-\frac{2}{3}\bar{u}\epsilon\left(c_{3}-10c_{4}\right)\right)} 
\nonumber \\    
    &-& \frac{\left(15c_3-3c_1+60c_4\right)+\epsilon\bar{u}\left(2c_3c_1-20c_4c_1-6c^2_3+60c_4c_3\right)}{\epsilon\left(c^2_3-20c_3c_4+100c^2_4\right)},
\end{eqnarray}
where the constant of integration $k_{o}$ can be determined by applying the boundary condition $\widetilde{k}(u_{-})=0$; this results in
\begin{eqnarray}
    k_{o}&=&\exp{\left(\frac{2}{3}u_{-}\epsilon\left(c_{3}-10c_{4}\right)\right)}
\nonumber \\    
    &+&\frac{\left(15c_3-3c_1+60c_4\right)+\epsilon u_{-}\left(2c_3c_1-20c_4c_1-6c^2_3+60c_4c_3\right)}{\epsilon\left(c^2_3-20c_3c_4+100c^2_4\right)}.
\end{eqnarray}
Therefore, the propagation velocity of the lead solitary wave edge of the DSW reads:
\begin{equation}
    s_{+}=(6u_{+}+\epsilon c_{1}u_{+}^{2})+\left(1+\epsilon c_{3}u_{+}\right)\widetilde{k}^2_{+} + \epsilon c_{4}\widetilde{k}^4_{+}. 
\end{equation}
Despite the effectiveness of the DSW fitting method, one should not assume {\it a 
priori} its universal applicability to dispersive hydrodynamic systems with 
non-convex dispersion.
In fact, this method can only provide accurate predictions for the macroscopic 
properties of RDSWs to a limited extent, as found in 
\cite{patkawahara,nemboreori,salehnem1}. Indeed, near the zero dispersion point 
($\omega_{kk}=0$), the stability and genuine nonlinearity of the DSW can be lost, 
leading to two possible scenarios. Firstly, there is the emission of a large resonant 
radiation wave propagating ahead of the leading solitary wave edge of the DSW, as 
observed in the RDSW, CDSW, and TDSW regimes. Secondly, the DSW may experience an 
internal collision of waves (implosion), as found in geophysical magma flows 
characterized by non-convex dispersion \cite{magmaimplosion}. In all these scenarios, 
the DSW deviates from the stable KdV-type form, which is a crucial assumption for the 
functionality of the DSW fitting method. The collapse of the conventional form of a 
DSW can be examined through the admissibility conditions that we discuss below.

\subsection{Admissibility conditions}\label{subsec:conditions}

In non-convex dispersive hydrodynamics, the assessment of admissibility conditions is 
crucial in determining the presence of a stable and classical form DSW. These 
conditions are linked to the genuine nonlinearity and hyperbolic nature of the Whitham 
modulation equations, which serve as the foundation for DSW solutions. Failure to meet 
these conditions leads to linear degeneracy or modulational instability, causing the 
breakdown of the standard DSW solution. The admissibility is expressed through partial 
derivatives of the trailing and leading edge velocities of the DSW with respect to the 
initial levels. Specifically, for a DSW characterized by initial levels $u_{-}$ behind 
and $u_{+}$ ahead, the following conditions must hold (see Ref.~\cite{hoefersci} for further discussion and detailed derivation):
\begin{equation}
    \frac{\partial s_{-}}{\partial u_{-}}\neq{0},\quad \frac{\partial s_{+}}{\partial u_{+}}\neq{0}, \quad  \frac{\partial s_{-}}{\partial u_{+}}\neq{0}, \quad 
    \frac{\partial s_{+}}{\partial u_{-}}\neq{0}. 
    \label{e:admis}
\end{equation}

The trailing and leading edge velocities of a \textit{stable} DSW exhibit monotonic behavior with respect to the variables $u_{-}$ and $u_{+}$ \cite{elreview}. Consequently, the fulfillment of the admissibility conditions is physically meaningful, as their violation would imply non-monotonic edge velocities, resulting in a multi-phase wavetrain. Thus, the admissibility conditions necessitate that the partial derivatives of the trailing edge velocity $s_{-}$ and leading edge velocity $s_{+}$ remain free of turning points concerning $u_{-}$ and $u_{+}$. The first two conditions in (\ref{e:admis}) stipulate that the Whitham modulation equations form a genuinely nonlinear system at the trailing and leading edges of the DSW, respectively. Failure of either condition precludes the possibility of a centered simple wave solution, which manifests a standard DSW, at the turning point. On the other hand, the last two conditions in (\ref{e:admis}) ensure that the Whitham modulation equations constitute a strictly hyperbolic system at the trailing and leading edges of the DSW, respectively. The breakdown of any of these criteria results in the loss of hyperbolicity in the Whitham modulation equations at the turning point, leading to compression and self-implosion within the DSW's interior structure, thus rendering it unstable \cite{hoefersci,salehnem1,xinwhitham}.

\subsection{Dispersive shock equal amplitude approximation}
\label{subsec:equal}

We have seen that the problem of finding DSW solutions without the Riemann invariant 
form can be addressed using the DSW fitting method. Another challenge that needs to be 
overcome arises when a dispersive hydrodynamic system is not hyperbolic outside the 
DSW region, meaning it is elliptic in the dispersionless limit, thus rendering the DSW 
unstable. In such cases, certain properties of the unstable DSW, such as its leading 
solitary wave edge amplitude $a_{+}$ and velocity $s_{+}$, and the number of solitary 
waves at time $t$, can be approximated using the DSW equal amplitude approximation, 
developed by Marchant and Smyth \cite{equalamp}.

In this method, the DSW is approximated by a series of $\mathcal{N}$ solitary waves, 
each of ``nearly'' uniform amplitude and evenly distributed, with a sharp decrease to 
the initial level behind at the trailing edge of the DSW. Initially, during the early 
phases of DSW development from an initial abrupt change, this approximation fails. 
However, as the DSW develops, it gradually becomes a more accurate 
approximation. This transition occurs because additional waves are generated within 
the DSW, causing the length of its leading edge ---which can be effectively modeled by 
solitary waves--- to increase. Consequently, as the DSW evolves, solitary waves 
increasingly govern its behavior. This assumption, or approximation, was justified for 
unstable DSWs in \cite{elfocusing}.

An additional interesting finding is that this method is also found to give a good approximation for stable DSWs. The key components of the DSW equal amplitude method include the existence of a solitary wave solution and mass and energy conservation laws. For simplicity and clarity, let us illustrate this method for KdV DSWs. Consider the KdV Riemann problem: equation (\ref{e:kdv}) subject to 
\begin{equation}
    u(x,0)= \left\{\begin{array}{cc}
         \Delta, \quad x<x_{0}, \\
          0, \quad ~x>x_{0}.
    \end{array}\right.
    \label{e:ic_eq}
\end{equation}
Here, $\Delta$ represents the magnitude of the jump, and $\Delta>0$ is required to generate a shock. This initial condition is equivalent to (\ref{ic_shock}) under certain scaling. The KdV mass and energy conservation laws are expressed as 
\begin{eqnarray}
&&  u_{t}+\left(3u^2+u_{xx}\right)_{x}= 0, \\
&&  \left(\frac{1}{2}u^{2}\right)_{t}+\left(2u^{3}+uu_{x}-\frac{1}{2}u^{2}_{x}\right)_{x}= 0.
\end{eqnarray}
Integrating the above conservation laws over the DSW domain, with $x$ ranging from $-\infty$ to $\infty$, yields the averaged equations:
\begin{equation}
 \mathcal{N}\frac{d}{dt}\left(\overline{\mathcal{U}_{\text{s}}}\right)=3\Delta^{2}\quad \text{and}\quad \mathcal{N}\frac{d}{dt}\left(\frac{1}{2}\overline{\mathcal{U}^2_{\text{s}}}\right)=2\Delta^{3}, \label{e:av_eqs}
\end{equation}
where the averaging rule used here is defined as 
\begin{equation}
    \overline{\mathcal{G}}=\int^{\infty}_{-\infty}g\,dx. 
    \label{e:equal_av_rule}
\end{equation}
For a single solitary wave solution (\ref{e:kdvsol}), we have
\begin{equation}
\overline{\mathcal{U}_{\text{s}}}=2\sqrt{2a_{+}}\quad \text{and}\quad \overline{\mathcal{U}^2_{\text{s}}}=\frac{4}{3}\sqrt{2a^{3}_{+}}. 
    \label{e:mass_soliton}
\end{equation}
Assuming no solitary waves exist at $t=0$ and utilizing (\ref{e:mass_soliton}), integrating the averaged equations (\ref{e:av_eqs}) and taking the ratio of them yields $a_{+}=2\Delta$. The leading solitary wave velocity can then be obtained by using the velocity-amplitude expression, $s_{+}=6\bar{u}+2a_{+}$, resulting in $s_{+}=4\Delta$ at the initial level ahead, as per condition (\ref{e:ic_eq}).

The same method can also approximate the number of solitary waves $\mathcal{N}(t)$ in the DSW at a given time $t$. Since the DSW is approximated by a train of equal amplitude solitary waves, each with amplitude $a_{+}=2\Delta$, then all the mass and energy of the initial condition (\ref{e:ic_eq}) is converted directly into solitary waves, and the total mass of the DSW at time $t$ is 
\begin{equation}
 \overline{\mathcal{U}}_{\text{DSW}}=\overline{\mathcal{U}}_{\text{s}}\mathcal{N}(t)=4\Delta \mathcal{N}(t).
\end{equation}
Therefore, the rate of change of the total mass leads to
\begin{equation}
    \frac{d}{dt}\left(\overline{\mathcal{U}}_\text{DSW}\right)=3\Delta^{2} \implies \mathcal{N}(t)=\frac{3}{4}\Delta^{3/2}t,
\end{equation}
with $\mathcal{N}(0)=0$. 

Since the eKdV equation (\ref{e:ekdv}) is a focal model in the present paper, we extend the application of the equal amplitude approximation to classical dispersive shocks governed by this model. 
The eKdV equation entails the mass conservation equation
\begin{equation}
    \frac{\partial}{\partial t}\mathcal{P}_{\text{m}} + \frac{\partial}{\partial x} \mathcal{Q}_{\text{m}}= 0, \label{e:ekdv_mass}
\end{equation}
and the energy conservation equation
\begin{equation}
    \frac{\partial}{\partial t}\mathcal{P}_{\text{e}} + \frac{\partial}{\partial x}\mathcal{Q}_{\text{e}} = 0, \label{e:ekdv_energy}
\end{equation}
where $\mathcal{P}_{\text{m}}$, $\mathcal{P}_{\text{e}}$, $\mathcal{Q}_{\text{m}}$, and $\mathcal{Q}_{\text{e}}$ represent the mass and energy densities and fluxes  defined, respectively, as follows:
\begin{eqnarray}
  \mathcal{P}_{\text{m}}  & = & u, \label{e:pm} \\
  \mathcal{P}_{\text{e}} & = & \frac{1}{2}u^{2} - \frac{1}{3} \epsilon
  \left(c_{3} - \frac{1}{2} c_{2}\right) u^{3},
  \label{e:pe} \\
  \mathcal{Q}_{\text{m}} & = & 3u^{2} + u_{xx} + \epsilon \left( \frac{1}{3} c_{1} u^{3} + c_{3}uu_{xx} + \frac{1}{2} \left( c_{2} - c_{3} \right) u_{x}^{2} + c_{4} u_{xxxx}
  \right), \label{e:qm} \\
  \mathcal{Q}_{\text{e}} & = & 2u^{3} + uu_{xx}
 - \frac{1}{2} u_{x}^{2} + \epsilon \left[ \frac{1}{4} c_{1} u^{4}
  + \frac{1}{2} c_{2}u^{2}u_{xx} + c_{4}uu_{xxxx}
 - c_{4}u_{x}u_{xxx} + \frac{1}{2} c_{4} u_{xx}^{2} \right.
 \nonumber \\
 & & \left. \mbox{} + \frac{3}{2} \left( \frac{1}{2} c_{2} - c_{3} \right) u^{4} \right]. \label{e:qe}
\end{eqnarray}
Note that while the mass conservation law is exact, the energy conservation law cannot be set in perfect derivative form, as the quantity $u^2/2$ is not conserved in the case of the eKdV equation. However, the energy conservation law can be expressed accurately at the order $\mathcal{O}(\epsilon)$, as shown in (\ref{e:pe}) and (\ref{e:qe}) ---see Ref.~\cite{salehekdv} for further details. 

We now integrate the eKdV conservation laws (\ref{e:ekdv_mass}) and (\ref{e:ekdv_energy}) over the unstable bore domain, where $x$ ranges from $-\infty$ to $\infty$. This yields the averaged equations:
\begin{eqnarray}
    \mathcal{N}\frac{d}{dt}\left(\overline{\mathcal{U}_{\text{s}}}\right) & = & 3\Delta^{2} +  \frac{1}{3}\epsilon c_{1}\Delta^{3}, \label{e:ekdv_mass_av} \\
    \mathcal{N}\frac{d}{dt}\left(\frac{1}{2}\overline{\mathcal{U}^2_{\text{s}}}  - \frac{1}{3} \epsilon
  \left(c_{3} - \frac{1}{2} c_{2}\right) \overline{\mathcal{U}^3_{\text{s}}}\right) &=& 2\Delta^{3} + \frac{1}{4}(c_{1}+3c_{2}-6c_{3})\Delta^{4},
  \label{e:ekdv_energy_av}
\end{eqnarray} 
with the averaging rule applied as stated above (\ref{e:equal_av_rule}). Unfortunately, the eKdV equation lacks integrability, it lacks an exact solitary wave solution due to the presence of higher order derivatives within the equation. Nevertheless, an approximate solution based on nonlocal perturbation theory can be derived \cite{tim_soliton}, expressed as
\begin{equation}
    u_{s} = { \left(a_{s}+\epsilon c_6a^2_{s}\right)\sech^{2}{w_{s}\theta_{s}} + \epsilon c_7 a^{2}_{s}\sech^{4}{w_{s}\theta_{s}} + \mathcal{O} (\epsilon^2) },
    \label{e:ekdV_soliton}
\end{equation}
with the phase $\theta_{s} = x - s_{+}t$, the inverse width $w_{s} = \sqrt{a_{s}/2}$ 
and the velocity 
\begin{equation}
    s_{+} = 2a_{s}+ 4\epsilon c_{4} a^2_{s}+\mathcal{O}(\epsilon^2),
    \label{e:tim_vel}
\end{equation}
for which the new constants $c_{6}$ and $c_{7}$ are
\begin{equation}
     c_6=- \frac{1}{6}c_1 + \frac{1}{6}c_2 + \frac{2}{3}c_{3} - 5c_4,\quad  c_7 = \frac{1}{12}c_1 - \frac{1}{4}c_2 - \frac{1}{2}c_{3} + \frac{15}{2}c_{4}.
     \label{e:c6_c7}
\end{equation}
Hence, when utilizing the expression (\ref{e:ekdV_soliton}), the averaged quantities in the densities described by equations (\ref{e:ekdv_mass_av}) and (\ref{e:ekdv_energy_av}) imply 
\begin{eqnarray}
    \overline{\mathcal{U}_{\text{s}}} & = & 2\sqrt{2a_{s}}+\epsilon\sqrt{2a^{3}_{s}}\left(2c_{6}+\frac{4}{3}c_{7}\right), \\
 \frac{1}{2}\overline{\mathcal{U}^2_{\text{s}}}  - \frac{1}{3} \epsilon
  \left(c_{3} - \frac{1}{2} c_{2}\right) \overline{\mathcal{U}^3_{\text{s}}} & = & \frac{2}{3}\sqrt{2a^{3}_{s}}+\epsilon\frac{4\sqrt{2}}{15}\sqrt{a^{5}_{s}}\left(5c_{6}+4c_{7}\right), 
\end{eqnarray}
Dividing the averaged equations (\ref{e:ekdv_mass_av}) and (\ref{e:ekdv_energy_av}) now results in the implicit equation
\begin{equation}
\frac{2\sqrt{2a_{s}}+\epsilon\sqrt{2a^{3}_{s}}\left(2c_{6}+\frac{4}{3}c_{7}\right)}{\frac{2}{3}\sqrt{2a^{3}_{s}}+\epsilon\frac{4\sqrt{2}}{15}\sqrt{a^{5}_{s}}\left(5c_{6}+4c_{7}\right)}=\frac{3\Delta^{2} +  \frac{1}{3}\epsilon c_{1}\Delta^{3}}{2\Delta^{3} + \frac{1}{4}(c_{1}+3c_{2}-6c_{3})\Delta^{4}}, 
\end{equation}
which determines the related-amplitude parameter $a_{s}$. Considering the higher-order 
solitary wave solution (\ref{e:ekdV_soliton}), we can infer that the total 
height/amplitude of the leading solitary wave edge of the DSW from the initial level 
ahead in (\ref{e:ic_eq}) is:
\begin{equation}
    a_{+}= a_{s} + \epsilon a^{2}_{s} \left(c_6+c_7 \right) 
    + \mathcal{O} (\epsilon^2),
\end{equation}
travelling at the velocity (\ref{e:tim_vel}). Furthermore, the number of solitary 
waves in the DSW can be approximated similarly to the approach used for KdV DSWs. The 
eKdV calculations yield in this case
\begin{equation}
 \mathcal{N}(t)=\left[\frac{3\Delta^{2} + \left(\epsilon c_{1}\Delta^{3}\right)/3}{2\sqrt{2a_{s}}+\epsilon\sqrt{2a^{3}_{s}}\left(2c_{6}+\frac{4}{3}c_{7}\right)}\right]t.
 \label{e:n_equal}
\end{equation}

The DSW equal amplitude method plays an important role in capturing specific 
properties of non-classical DSWs governed by non-convex dispersive hydrodynamic 
systems, such as the eKdV equation (\ref{e:ekdv}). It has proven effective in 
predicting the leading solitary wave amplitude and velocity within the highly unstable 
undular bore portion in the CDSW regime, depicted in Fig.~\ref{f:dswsregimes}(c). 
As for the portion that is featured by a resonant radiation propagating ahead of the 
unstable bore, the utilization of the concept of Whitham shocks becomes necessary to 
analyze it.

\subsection{Whitham shocks: modulation theory jump conditions}\label{subsec:jumps}

After having presented the key methods for studying stable bores in RDSW regimes 
and for examining unstable bores in CDSW regimes, we can 
proceed with the analysis of the attached resonant radiation in the CDSW and TDSW 
regimes. Given the nonlinearity of our underlying problem and the relatively small 
amplitude of the resonant radiation compared to the leading amplitudes of the bore 
behind itself, we can approximate the resonant wavetrain using a weakly nonlinear wave 
expansion solution, known as a Stokes wave. The latter takes the form: 
\begin{eqnarray}
u_{r} & = & \bar{u}_{r}+a_{r}\cos(\theta_{r})+a^{2}_{r}u_{2}\cos(2\theta_{r})+\mathcal{O}(a^3_r), \label{e:stokes_gen_u} \\
\omega_{r}&=&\omega_{0}+a_{r}\omega_{1}+a^{2}_{r}\omega_{2}+\mathcal{O}(a^3_r).  \label{e:stokes_gen_w}
\end{eqnarray}
Here, $\bar{u}_{r}$ denotes the mean level of the resonant wavetrain, the parameter 
$a{r}$ represents the resonant amplitude, while $\omega_{r}$ and $k_{r}$ stand for the 
Stokes frequency and wavenumber, respectively, with $\theta_{r}=k_{r}x-\omega_{r}t$. 
The Stokes wave coefficients $u_{2}$, $\omega_{0}$, $\omega_{1}$, and $\omega_{2}$ can 
be obtained by substituting the expansions (\ref{e:stokes_gen_u}) and 
(\ref{e:stokes_gen_w}) into the corresponding dispersive hydrodynamic system and 
working out the resulting equations up to the asymptotic order 
$\mathcal{O}{(a^{2}_{r})}$.

The determination of the Stokes wave parameters $(k_{r},\bar{u}_{r},a_{r})$ hinges on connecting the resonant wavetrain ahead with the wavetrain behind via a Whitham shock. A Whitham shock is a moving discontinuous shock solution to Whitham modulation equations in conservation law form. The concept of the Whitham shock, first introduced by Sprenger and Hoefer \cite{patjump}, can be seen as the dispersive equivalent of the Rankine-Hugoniot jump conditions for classical gas dynamics. Initially, when Whitham developed modulation theory, he speculated on the role of shocks for which the modulation equations were hyperbolic, indicating that the underlying periodic wavetrain is stable. However, he did not explore the topic extensively \cite{whitham}. To implement the procedure of the Whitham shock, conservation laws (\ref{e:mod_system}) need to be set as the Rankine-Hugoniot jump conditions:
\begin{equation}
    -U_{s}\llbracket \overline{\mathcal{P}_{i}}(\underline{\alpha}) \rrbracket + \llbracket \overline{\mathcal{Q}_{i}}(\underline{\alpha}) \rrbracket =0.
\end{equation}
Here, $U_{s}$ denotes the Whitham shock velocity and the bracket $\llbracket . \rrbracket$ represents the difference between the left (-) and right (+) jump quantities in the shock. The modulation variables $ \underline{\alpha}=(\alpha_{1},\alpha_{2},\ldots,\alpha_{n})$ are now shock solutions to the averaged conservation laws. These take the form of a traveling discontinuity
\begin{equation}
    \underline{\alpha}(x,t)= \left\{ \begin{array}{cc}
         \underline{\alpha}_{-},\quad x<U_{s}t, \\
          \underline{\alpha}_{+},\quad ~x>U_{s}t.
    \end{array}\right.
\end{equation}

In the context of the eKdV problem (\ref{e:ekdv}), which is a focal point of 
this work, the Stokes wave coefficients can be found upon substituting (\ref{e:stokes_gen_u}) and (\ref{e:stokes_gen_w}) into the eKdV equation (\ref{e:ekdv}) and eliminating secular terms \cite{salehekdv}; their form is:
\begin{eqnarray}
& & \omega_{0} = (6\bar{u}_{r}+\epsilon c_{1}\bar{u}^{2}_{r})k_{r}-\left(1+\epsilon c_{3}\bar{u}_{r}\right)k^3_{r} + \epsilon c_{4} k^5_{r}, \quad \omega_{1}=0, \label{e:omega0} \\
& &  \omega_{2} = \frac{36+24\epsilon c_1\bar{u}_{r} - \epsilon \left(48c_3-6c_1-6c_2\right)k^{2}_{r}}{24k_{r} - \epsilon\left(120c_4k^3_{r}-24c_3\bar{u}_{r}k_{r}\right)}+\mathcal{O}\left(\epsilon^2\right)
\nonumber \\
& & ~~~= \frac{3}{2k_{r}}+\epsilon\left[\frac{1}{4}\left( c_{1}+c_2 -8c_{3}+ 30 c_{4}\right)k_{r} + \left( c_{1} - \frac{3}{2}c_{3} \right) \frac{\bar{u}_{r}}{k_{r}}\right]+\mathcal{O}\left(\epsilon^2\right) \label{e:omega2}
\end{eqnarray}
and
\begin{eqnarray}
 u_{2} &=& \frac{6+2\epsilon c_{1}\bar{u}_{r}-\epsilon (c_{2}+c_3)k^2_{r}}{12k^2_{r} + 12\epsilon\left(c_{3}\bar{u}_{r} - 5c_4k^2_{r}\right)k^{2}_{r}}  
\nonumber \\ 
&=&
 \frac{1}{2k^2_{r}} - \frac{1}{12}\epsilon\left[ c_2+c_3 - 30 c_{4} - 2\left(c_{1} - 3c_{3} \right) \frac{\bar{u}_{r}}{k^2_{r}}\right] + \mathcal{O}\left(\epsilon^2\right). \label{e:u2}
\end{eqnarray}
Then, to compute the Whitham shock for the eKdV Riemann problem, we require the 
conservation laws of mass and energy given by (\ref{e:ekdv_mass}) and (\ref{e:ekdv_energy}).

\subsubsection{CDSW regime}

In the eKdV CDSW regime, we conceptualize the DSW form transition as a rapid 
discontinuity that links the unstable bore behind with the unstable resonant wavetrain 
through a Whitham shock. The unstable bore can be effectively approximated, on 
average, as a series of solitary waves of equal amplitude, while the unstable resonant 
wavetrain can be similarly approximated, on average, as a Stokes wave 
(\ref{e:stokes_gen_u}). Utilizing the DSW equal amplitude method is vital for modeling 
the unstable bore portion, as discussed previously in Subsection \ref{subsec:equal}. 
To determine the Whitham shock for the CDSW, we must employ the mass 
(\ref{e:ekdv_mass}) and energy (\ref{e:ekdv_energy}) conservation laws and set them in 
the form of Rankine-Hugoniot jump conditions, 
\begin{eqnarray}
&&    -U_{s}\left(\mathcal{\overline{P}}_{\text{m,bore}}-\mathcal{\overline{P}}_{\text{m,Stokes}}\right)+\left(\mathcal{\overline{Q}}_{\text{m,bore}}-\mathcal{\overline{Q}}_{\text{m,Stokes}}\right)=0,
    \label{e:jump_mass} \\
&&    -U_{s}\left(\mathcal{\overline{P}}_{\text{e,bore}}-\mathcal{\overline{P}}_{\text{e,Stokes}}\right)+\left(\mathcal{\overline{Q}}_{\text{e,bore}}-\mathcal{\overline{Q}}_{\text{e,Stokes}}\right)=0,
    \label{e:jump_energy}
\end{eqnarray}
subject to the jumps in the modulation variables 
\begin{equation}
   \bar{u}(x,t)= \left\{ \begin{array}{cc}
         \bar{u}_{s},~~x<U_{s}t,\\
         \bar{u}_{r},~~x>U_{s}t,
    \end{array}\right. \quad 
    a(x,t)= \left\{ \begin{array}{cc}
         a_{s},~~x<U_{s}t,\\
         a_{r},~~x>U_{s}t,
    \end{array}\right. \quad 
    k(x,t)= \left\{ \begin{array}{cc}
         k_{s},~~x<U_{s}t,\\
         k_{r},~~x>U_{s}t.
    \end{array}\right.
    \label{e:cdsw_jumps}
\end{equation}
In this context, the averaging rule over the Stokes wave is defined as:
\begin{equation}
    \overline{\mathcal{G}}_{\text{Stokes}}=\frac{1}{2\pi}\int^{2\pi}_{0}\mathcal{G}\left(u,u_{\theta},u_{\theta\theta},\ldots\right)d\theta,
\end{equation}
where $u$ is given by the Stokes wave expansion (\ref{e:stokes_gen_u}), with the coefficients (\ref{e:omega0})--(\ref{e:u2}). On the other hand, the averaging rule over the unstable bore is defined as
\begin{equation}
    \overline{\mathcal{G}}_{\text{bore}}=\int^{\infty}_{-\infty}\mathcal{G}\left(u,u_{\theta},u_{\theta\theta},\ldots\right)d\theta,
    \label{e:sol_av}
\end{equation}
where $u$ is a solitary wave solution to the eKdV equation (\ref{e:ekdV_soliton}). However, in this case, the background of the solitary wave needs to be incorporated into the solution expression (\ref{e:ekdV_soliton}), namely, 
\begin{equation}
    u_{s} = { \bar{u}_{s} + \left(a_{s}+\epsilon c_6a^2_{s}\right)\sech^{2}{w_{s}\theta_{s}} + \epsilon c_7 a^{2}_{s}\sech^{4}{w_{s}\theta_{s}} + \mathcal{O}(\epsilon^2) }.
\end{equation}
The leading solitary wave edge background is crucially involved in the Whitham jump 
system (\ref{e:cdsw_jumps}). Indeed, the resonant radiation smoothly elevates the 
initial level ahead, $u_{+}$, to the mean level of the resonant wavetrain, 
$\bar{u}_{r}$. This mean level then undergoes a discontinuous jump to the mean level of the unstable undular bore, $\bar{u}_{s}$, via the Whitham shock, which connects the resonant wavetrain to the undular bore. This theoretical and numerical justification is provided in \cite{salehekdv}. 

The eKdV Whitham modulation jump system involves several modulation wave parameters, 
namely $U_{s}$, $\bar{u}_{s}$, $a_{s}$, $a_{r}$, $k_{r}$, $k_{s}$, and $\bar{u}_{r}$. 
With seven unknown wave parameters, we require seven equations. However, the 
wavenumber $k_{s}$ in the solitary wave limit is zero, as the solitary wave has an 
infinite wavelength \cite{whitham}. Moreover, $\bar{u}_{r}$ is found to be very close 
to the initial level ahead of the shock $u_{+}$, so that $\bar{u}_{r}=u_{+}$. 
Thus, the number of the unknowns are reduced to five. In addition to the Whitham 
jumps (\ref{e:jump_mass}) and (\ref{e:jump_energy}), the DSW equal amplitude 
approximation method, applied to the eKdV equation, leads to an implicit equation for 
the solitary wave related-amplitude parameter $a_{s}$ and another equation for the 
velocity (\ref{e:tim_vel}); $U_{s}=s_{+}$. However, due to the presence of resonant radiation ---approximated by a Stokes wave--- propagating ahead of the bore, this effect must be incorporated into the analysis. Consequently, the implicit relation obtained from the DSW equal amplitude method is modified to
\begin{equation}
\frac{2\sqrt{2a_{s}}+\epsilon\sqrt{2a^{3}_{s}}\left(2c_{6}+\frac{4}{3}c_{7}\right)}{\frac{2}{3}\sqrt{2a^{3}_{s}}+\epsilon\frac{4\sqrt{2}}{15}\sqrt{a^{5}_{s}}\left(5c_{6}+4c_{7}\right)}=\frac{\left(3\Delta^{2} +  \frac{1}{3}\epsilon c_{1}\Delta^{3}\right)-\mathcal{\overline{Q}}_{\text{m,Stokes}}}{\left(2\Delta^{3} + \frac{1}{4}(c_{1}+3c_{2}-6c_{3})\Delta^{4}\right)-\mathcal{\overline{Q}}_{\text{e,Stokes}}}.
\end{equation}
This leaves us with four equations. The 
last equation required to close the system is the resonance condition, which states 
that the lead solitary wave velocity of the unstable bore, essentially the Whitham 
shock velocity $U_{s}$, propagates at the same phase velocity as the resonant 
radiation ahead. This leads to the equation:
\begin{equation}
    U_{s}=\frac{\omega_r}{k_r}=\frac{\omega_{0}+a^2_{r}\omega_{2}}{k_{r}},
    \label{e:resonance_condition}
\end{equation}
where we have used the dispersion relation 
(\ref{e:stokes_gen_w}). Note that the resonance condition here is equivalent to the modulation jump for the conservation of waves (\ref{e:closure}), which can be expressed as: 
\begin{equation}
    -U_{s}\left(k_s - k_r\right)+\left(\omega_s - \omega_r\right)=0,
\end{equation}
with $k_{s}=\omega_{s}=0$ at the lead solitary wave edge. This then finishes our 
analysis for the CDSW regime through the approach of Whitham shock. 
For full comparisons against numerical simulations for the CDSW regime and detailed discussions, 
see Ref.~\cite{salehekdv}.

\subsubsection{TDSW regime}

The TDSW regime, governed by the eKdV equation (\ref{e:ekdv}), can be modeled 
similarly. In this case, the unstable bore is replaced by a negative polarity 
oscillatory solitary wave of very small amplitude, which connects to the resonant 
radiation ahead. The resonant wavetrain now appears almost uniform and stable due to 
the presence of a partial DSW that elevates the resonant mean level $\bar{u}_{r}$, 
resulting in stability. However, $\bar{u}_{r}$ is still found to be very close to 
$u_{+}$ in this regime, so we can approximately set $\bar{u}_{r}=u_{+}$. For a full 
calculation of the resonant mean level $\bar{u}{r}$, the partial DSW structure needs 
to be calculated using weakly nonlinear modulation theory. A detailed study on this 
can be found in Refs.~\cite{pat,kdvbound1,kdvbound2}. The disparity in height between 
the negative oscillatory solitary wave's amplitude and the initial level behind 
$u_{-}$ is insignificant, allowing us to set the solitary wave amplitude parameter 
to zero ($a_{s}=0$). The modulation jumps in the wavenumber, mean level, and amplitude 
become:
\begin{equation}
   \bar{u}(x,t)= \left\{ \begin{array}{cc}
         u_{-},~~x<U_{s}t\\
         u_{+},~~x>U_{s}t,
    \end{array}\right. \quad 
    a(x,t)= \left\{ \begin{array}{cc}
         0,~~x<U_{s}t\\
         a_{r},~~x>U_{s}t,
    \end{array}\right.\quad 
    k(x,t)= \left\{ \begin{array}{cc}
         0,~~x<U_{s}t\\
         k_{r},~~x>U_{s}t,
    \end{array}\right.
\end{equation}
and the Whitham jump conditions are
\begin{eqnarray}
&&    -U_{s}\left(\mathcal{\overline{P}}_{\text{m},u_{-}}-\mathcal{\overline{P}}_{\text{m,Stokes}}\right)+\left(\mathcal{\overline{Q}}_{\text{m},u_{-}}-\mathcal{\overline{Q}}_{\text{m,Stokes}}\right)=0,
    \label{e:jump_mass_tdsw}
    \\
&&    -U_{s}\left(\mathcal{\overline{P}}_{\text{e},u_{-}}-\mathcal{\overline{P}}_{\text{e,Stokes}}\right)+\left(\mathcal{\overline{Q}}_{\text{e},u_{-}}-\mathcal{\overline{Q}}_{\text{e,Stokes}}\right)=0,
    \label{e:jump_energy_tdsw}
\end{eqnarray}
where we have used the notation $\overline{\mathcal{G}}_{u_{-}}=\overline{\mathcal{G}}|_{\bar{u}=u_{-}}$. The Whitham shock system can now be closed by the resonance condition (\ref{e:resonance_condition}), and the entire system of equations determines $U_{s}$, $a_{r}$ and $k_{r}$. This concludes the analysis of the TDSW regime using Whitham shocks.

\begin{figure}
    \centering
\includegraphics[width=0.49\textwidth]{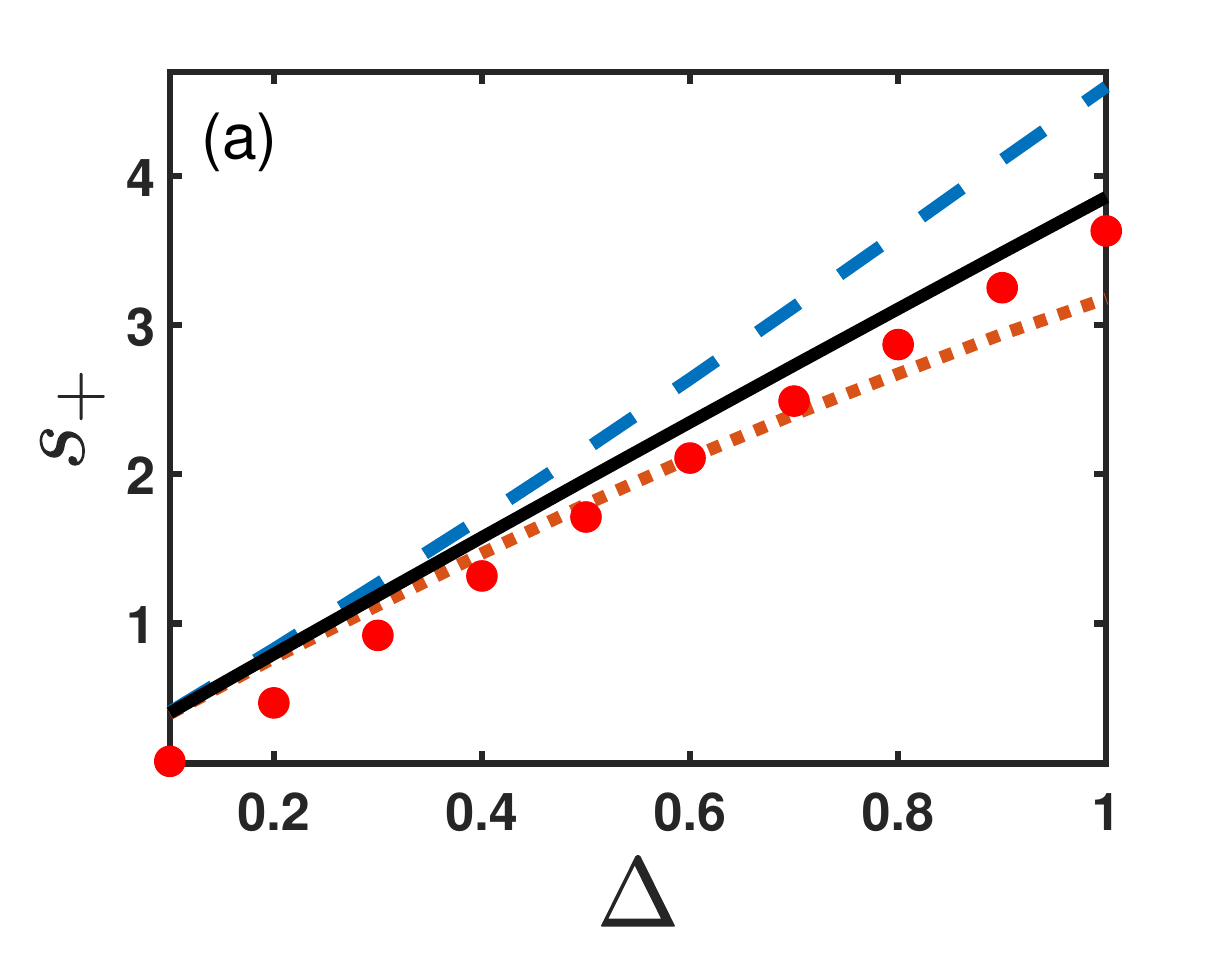}
\includegraphics[width=0.49\textwidth]{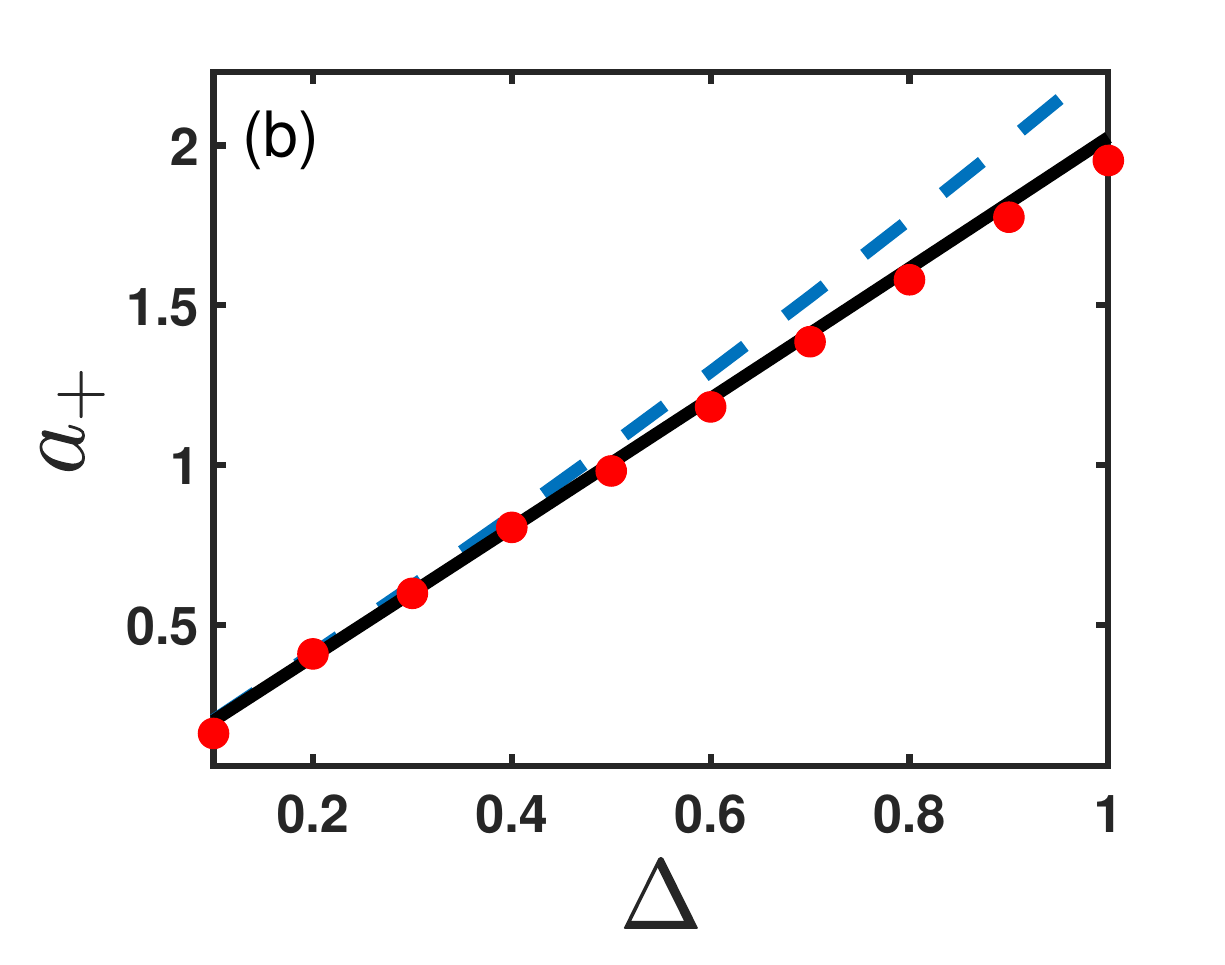}
\includegraphics[width=0.49\textwidth]{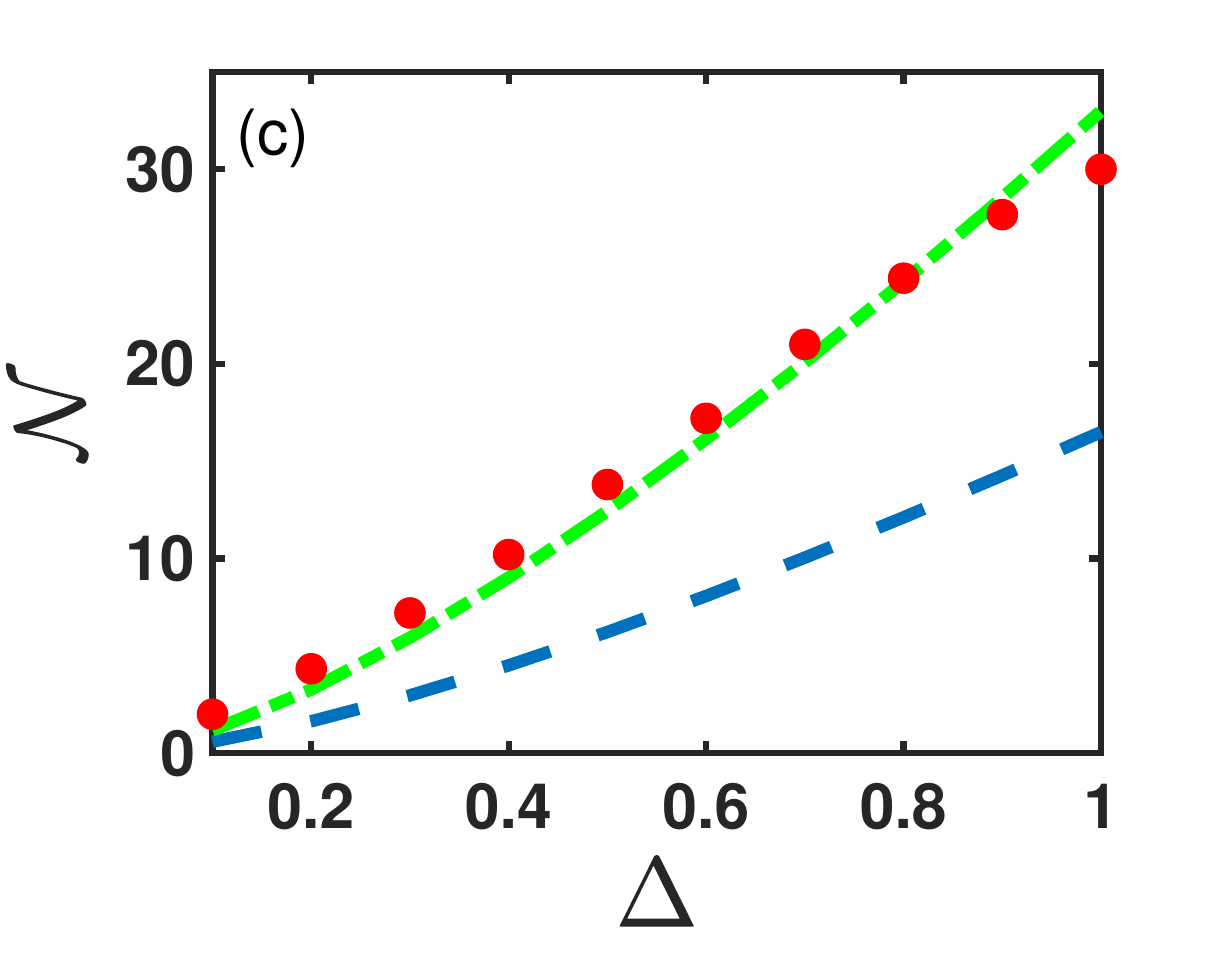} \caption{Wave parameter comparisons in the case of shallow water RDSW regime. Red dots: numerical solutions; black (solid) line: higher order modulation solutions; orange (dotted) line: DSW fitting solutions and blue (dashed) line: DSW equal amplitude solutions. (a) Lead solitray wave edge velocity comparisons, (b) Lead solitray wave edge amplitude comparisons, and (c) Number of solitary waves encapsulated in the bore envelope, where the green (dashed-dotted) line marks twice the DSW equal amplitude predictions. In these plots, $t=25$, $\epsilon=0.15$, $c_{1}=-3/2$, $c_{2}=23/4$, $c_{3}=5/2$, $c_{4}=19/40$. (Color version online).}
     \label{f:comps_rdsw}
\end{figure}

\section{Numerical results and discussion}

Next, we proceed with comparing the theoretical predictions of the previous Section 
with results of direct numerical simulations. Notice that below 
we will not present and compare  
Whitham shock theoretical solutions for the eKdV CDSWs with numerical solutions, 
since relevant extensive comparisons have been reported in Ref.~\cite{salehekdv}. 

As far as our numerical approach is concerned, we tackle the eKdV 
Riemann problem, defined by Eq.~(\ref{e:ekdv}) and the initial 
discontinuity~(\ref{ic_shock}), using the pseudo-spectral method pioneered by 
Whitham and Fornberg \cite{fornberg}. This method involves discretization of 
the spatial domain employing the Fast Fourier Transform (FFT) technique. 
The result is an ODE with respect to the time independent variable $t$ for 
the Fourier transform of the dependent variable $u$. Instead of adopting 
Whitham and Fornberg's centered-difference scheme for integrating the time domain, 
we opt for the 4th-order Runge-Kutta method (RK4) due to its enhanced accuracy. 
Additionally, to ensure stability at high frequencies and eliminate stiff terms 
causing delays in achieving stability at extremely small resolutions, we implement the 
method of integrating factor ---see Ref.~\cite{trefethen} for details.
Since Fourier methods necessitates a periodic domain, we smooth the initial 
discontinuity (\ref{ic_shock}) upon using a hyperbolic tangent profile, akin to an 
initial well of the form:
\begin{equation}
    u(x,0)=u_{+}+\frac{1}{2}\left(u_{-}-u_{+}\right)\left(\tanh\left(\frac{x+x_0}{W}\right)-\tanh\left(\frac{x}{W}\right)\right), 
\end{equation}
where $W$ denotes the width and $x_{0}$ marks the coordinate location at which the 
initial jump descends to the initial level ahead $u_{+}$. Through numerical 
simulations, it is found that $W=1$ yields satisfactory results, and 
$\Delta{t}=\mathcal{O}\left(10^{-4}\right)$ is adequate to achieve stability 
with the shallow water wave coefficients~(\ref{e:shallowcoef}). The numerical 
solutions obtained through this scheme will now be utilized to validate the 
analytical solutions of the eKdV RDSWs and TDSWs.

\begin{figure}[!ht]
    \centering
    \includegraphics[width=0.49\textwidth]{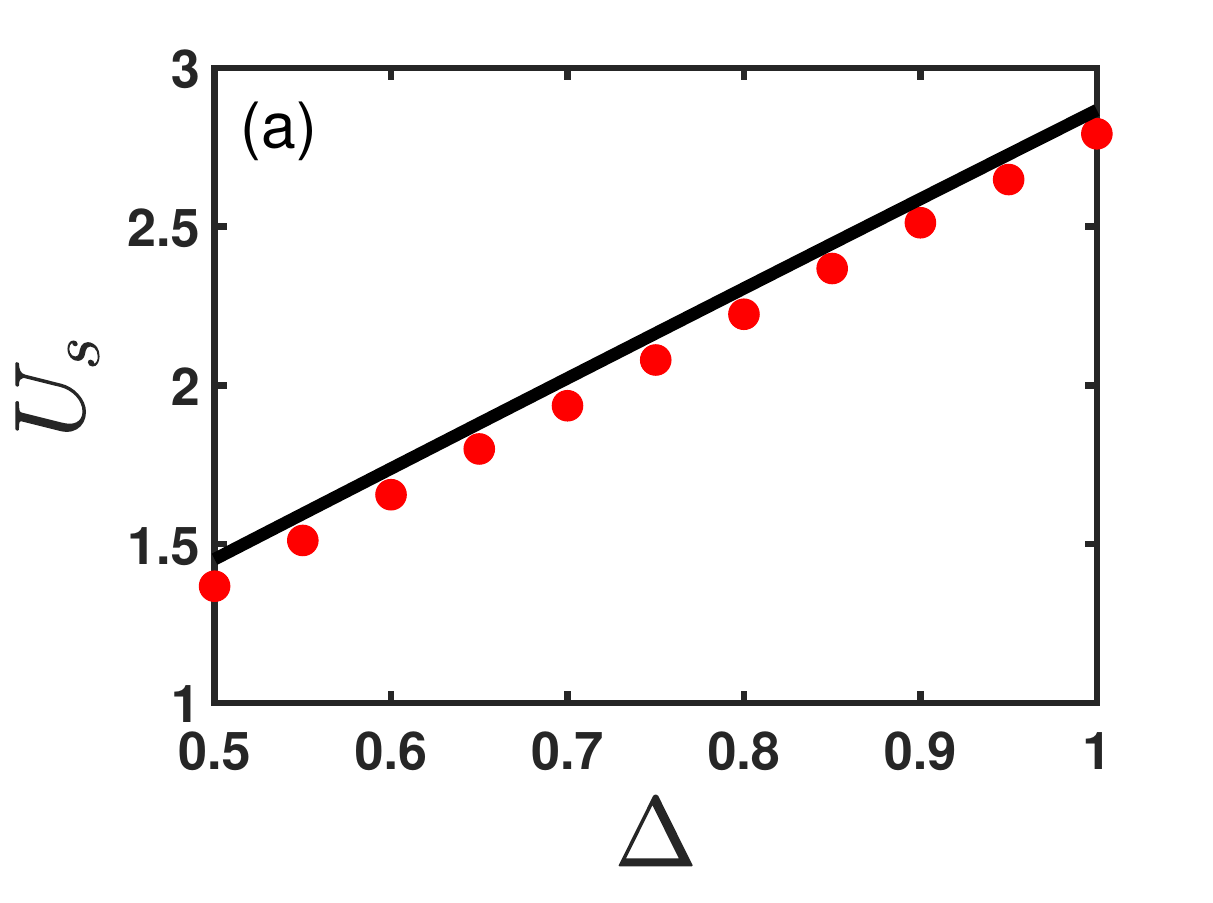}
    \includegraphics[width=0.49\textwidth]{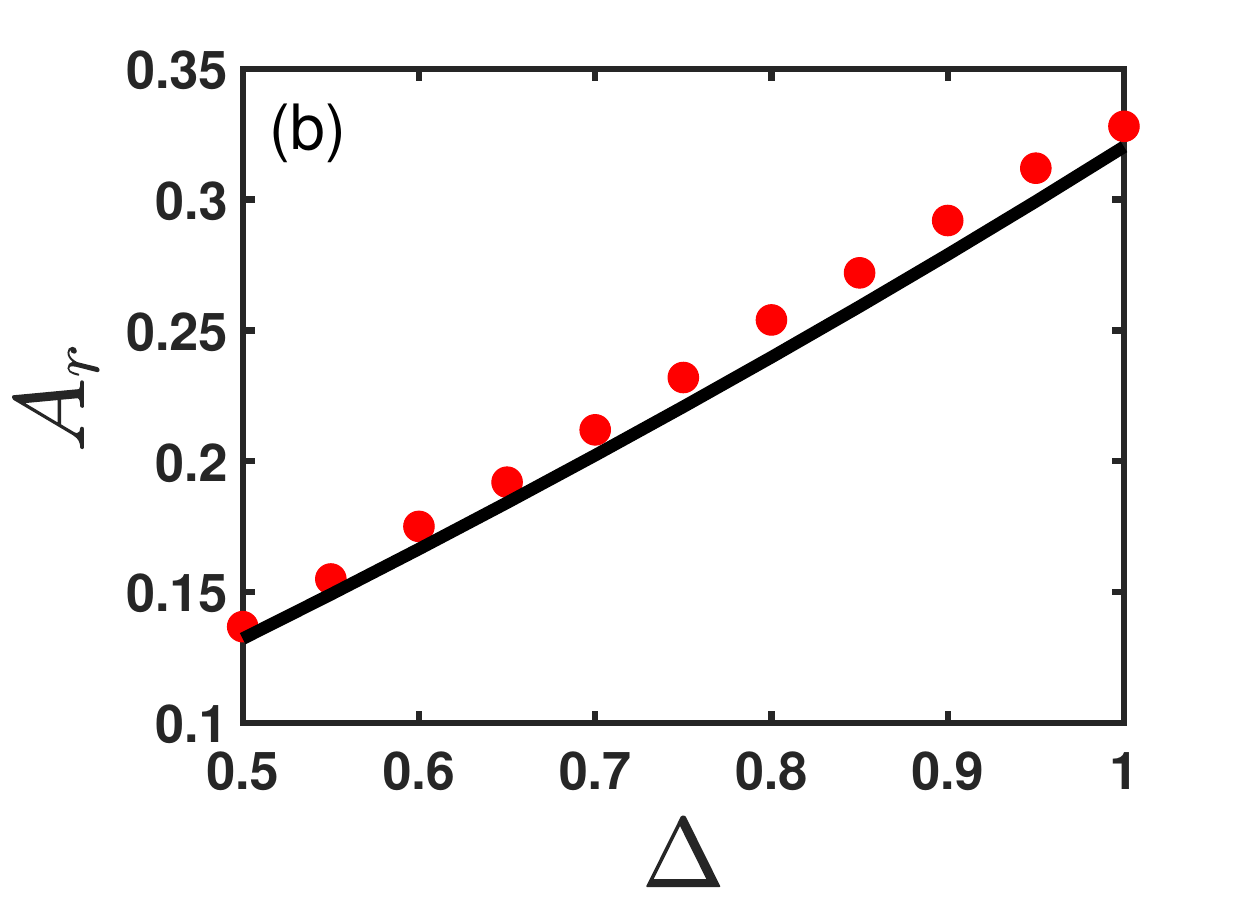}
    \includegraphics[width=0.49\textwidth]{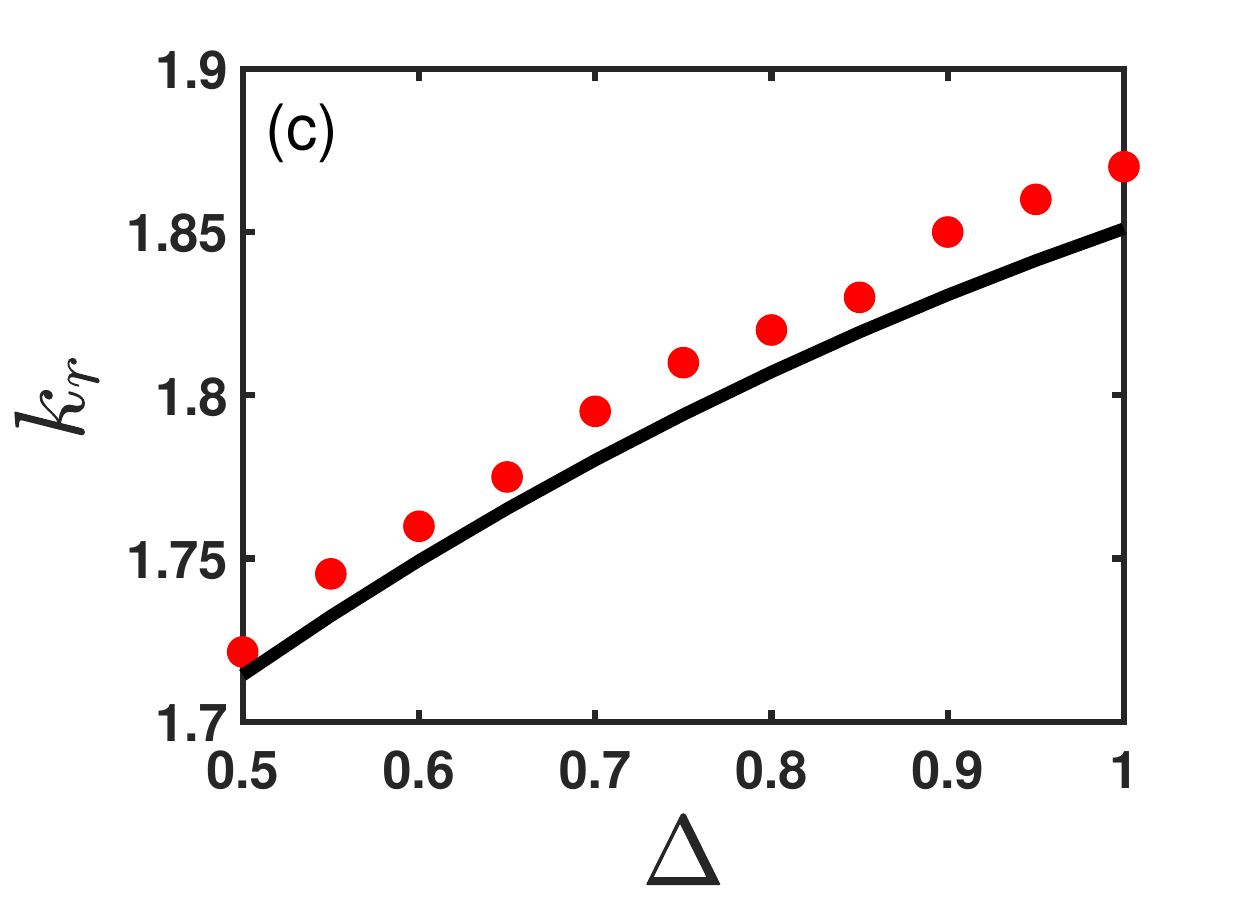} \caption{Resonant wave parameter comparisons in the case of the eKdV TDSW regime for various initial jumps. (a) Whitham shock velocity comparisons, (b) resonant wave amplitude comparisons, and (c) resonant wavenumber comparisons. Red dots: numerical solutions; black (solid) line: Whitham shock solution predictions. Here, $t=25$, $\epsilon=0.15$, $c_{1}=c_{2}=c_{3}=0.3$, $c_{4}=3.0$. (Color version online).}
     \label{f:comps_tdsw}
\end{figure}

Figures \ref{f:comps_rdsw}(a) and \ref{f:comps_rdsw}(b) present a comparative analysis between full 
numerical solutions of the eKdV equation and several theoretical approaches including 
higher order modulation theory, the DSW fitting method, and the equal amplitude 
approximation theory. These comparisons focus on assessing the leading solitonic edge 
velocity and the amplitude of the leading solitary wave edge from the stationary state 
$u_{+}$. To maintain consistency, we use  
$\epsilon=0.15$ to uphold the characteristic structure of a RDSW regime. Furthermore, 
we set $u_{-}=1$ and vary the steady state ahead $u_{+}$. The determination of the 
higher order coefficients $c_{i}\,(i=1,2,3,4)$ in equation (\ref{e:ekdv}) is based on 
the shallow water wave constants (\ref{e:shallowcoef}). Overall, the observed 
alignment between theoretical predictions and numerical solutions, particularly across 
a diverse range of initial jumps associated with the RDSW regime, is excellent. It is 
evident that higher order modulation theory serves as the best approximation among 
other methods, with the agreement being nearly perfect. In contrast, the other methods 
demonstrate less accurate approximate solutions, especially when the initial jump 
value becomes large. This discrepancy is expected, however. As $\Delta$ increases, the 
resonant radiation amplitude becomes relatively larger, leading to a decay in the RDSW 
structure, and the validity of the utilized methods diminishes. Furthermore, it can be 
seen that the DSW fitting method offers better approximate solutions compared to those 
derived from the DSW equal amplitude approximation theory. In Fig.~\ref{f:comps_rdsw}(a), the 
maximum error in the fitting method is 4\%, whereas it is 9\% in the latter method. 
The discrepancy stems from the fact that the equal amplitude method assumes that the 
RDSW is primarily composed of nearly-equal amplitude solitary waves, sharply 
descending at the trailing edge of the bore, as mentioned in 
Subsection \ref{subsec:equal}. However, this assumption does not precisely reflect the 
reality captured in numerical simulations. The key assumption of the equal amplitude 
method tends to hold stronger for unstable DSW regimes. Despite this limitation, 
however, the overall agreement of the equal amplitude approximation across the RDSW 
regime remains quite satisfactory. In Fig.~\ref{f:comps_rdsw}(b), the theoretical solutions 
for lead solitonic amplitude from the DSW fitting method are absent. This absence is 
due to the lack of a velocity-amplitude relation for the eKdV equation.

\begin{figure}[!ht]
    \centering
    \includegraphics[width=1.0\textwidth]{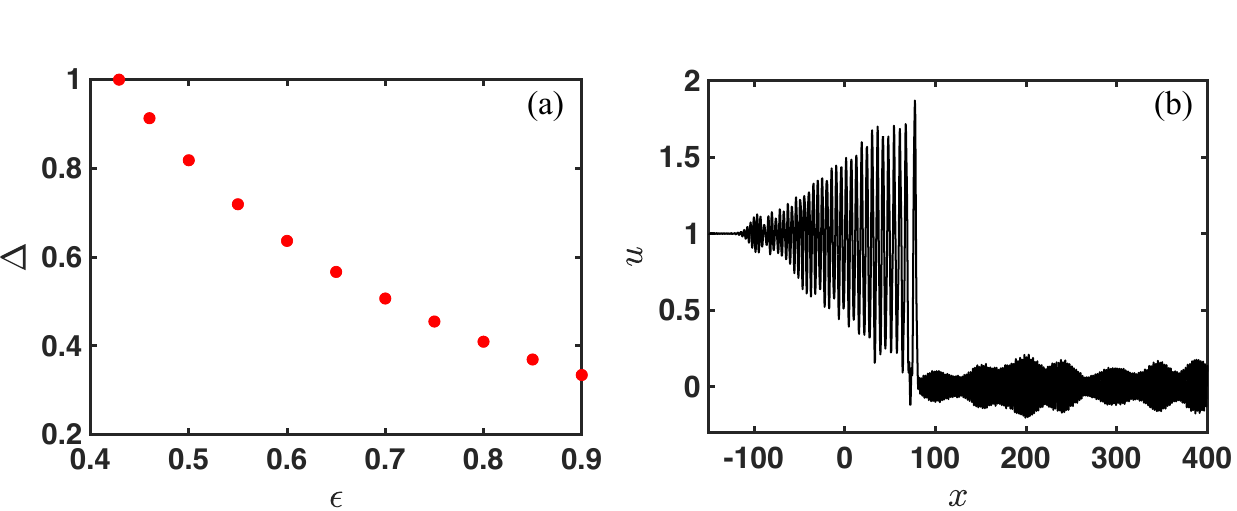}
    \caption{Failure of admissibility condition $\partial_{u_{+}}s_{-}\neq{0}$ and a breakdown of shallow water dispersive shock structure. (a) Red dots marking the turning points against varying the nonlinearity parameter $\epsilon$, (b) a snapshot of shallow water undular bore losing its modulational stability at the initial jump $\Delta=0.999$ and the nonlinearity parameter value $\epsilon=0.429$. Here, $t=25$, $c_{1}=-3/2$, $c_{2}=23/4$, $c_{3}=5/2$, $c_{4}=19/40$. (Color version online).}
     \label{f:instability}
\end{figure}

Figure~\ref{f:comps_rdsw}(c) illustrates comparisons between the number of lead solitary 
waves predicted in the bore by the DSW equal amplitude approximation and the actual 
number of lead solitary waves computed from numerical solutions. It can be seen that 
the agreement is poor. The analytical prediction relies on 
the assumption that nearly fifty percent of the numerical mass of the DSW produced 
from the initial jump is transformed into solitary waves. Indeed, doubling the 
predicted values $\mathcal{N}$ yields excellent agreement with numerics. 
Once again, the disparity in the comparison plot is expected, as the actual numerical 
undular bores in the RDSW regime do not possess precisely uniform amplitude lead 
solitary waves, a pivotal assumption underlying the approximation method.

Figures \ref{f:comps_tdsw}(a)--\ref{f:comps_tdsw}(c) depict comparisons between Whitham shock solutions for the eKdV TDSW regime and exact numerical solutions, with general non-zero higher order coefficients $c_{i}\,(i=1,2,3,4)$. These comparisons encompass the Whitham shock velocity $U_{s}$, as well as the wavenumber $k_{r}$ and amplitude $A_{r}=a_{r}+a^{2}_{r}u_{2}$ for the resonant wavetrain ahead. The resonant wavenumber can be determined as $k_{r}=2\pi/\lambda_{r}$, where $\lambda_{r}$ represents the averaged wavelength. It is found that averaging over 20 wave crests is sufficient. Given the near-uniform nature of the attached resonant wave with minor modulations, the resonant amplitude is numerically determined by averaging across the resonance region up to its leading edge. Overall, it can be seen that the agreement between theory and numerical simulations is nearly perfect across the TDSW regime.

\begin{figure}
    \centering
    \includegraphics[width=1.0\textwidth]{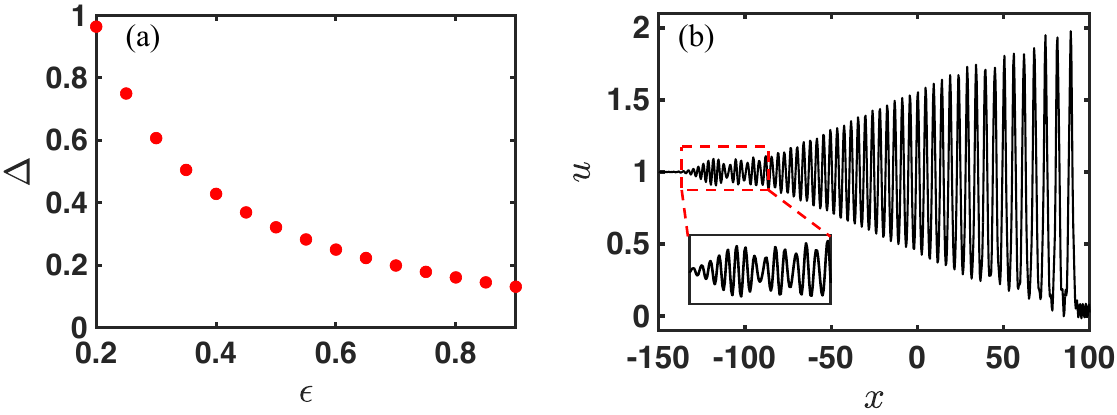}
    \caption{Failure of admissibility condition $\partial_{u_{-}}s_{-}\neq{0}$ and loss of monotonic velocity (linear degeneracy) at the trailing edge of the shallow water dispersive shock. (a) Red dots marking the turning points against varying the nonlinearity parameter $\epsilon$, (b) a snapshot of shallow water undular bore revealing non-classical wave modulations (multi-speed waves) at the trailing edge of the bore, with the initial jump $\Delta=0.965$ and the nonlinearity parameter value $\epsilon=0.2$. Here, $t=25$, $c_{1}=-3/2$, $c_{2}=23/4$, $c_{3}=5/2$, $c_{4}=19/40$. (Color version online).}
     \label{f:lindeg}
\end{figure}

Figures \ref{f:instability} and \ref{f:lindeg} depict the loss of genuine nonlinearity and modulational stability in shallow water dispersive shocks, as deduced from the DSW admissibility conditions. In these graphical representations, we set the initial steady state $u_{-}=1$ and investigate the admissibility criteria (\ref{e:admis}), where $s_{+}$ and $s_{-}$ are derived from solutions in higher modulation theory, which are equations (\ref{e:extended_leading}) and (\ref{e:extended_trailing}). It is noteworthy that while one could opt to utilize the leading and trailing edge velocities from alternative methods discussed previously, the solutions derived from higher order modulation theory are employed here due to their better accuracy. It has been observed that the admissibility criteria $\partial_{u_{+}}s_{+}\neq{0}$ and $\partial_{u_{-}}s_{+}\neq{0}$ do not exhibit turning points within the shock range $0<u_{+}<u_{-}$. Consequently, theoretical predictions of linear degeneracy at the leading edge and modulational instability at the trailing edge of the DSW are not supported. However, the other admissibility criteria, namely $\partial_{u_{+}}s_{-}\neq{0}$ and $\partial_{u_{-}}s_{-}\neq{0}$, do possess turning points within the same shock range $0<u_{+}<u_{-}$. The derivative $\partial_{u_{+}}s_{-}$ begins to vanish at $\epsilon=0.429$, corresponding to $u_{+}=0.001$ (thus $\Delta=0.999$), while $\partial_{u_{-}}s_{-}$ starts to vanish at $\epsilon=0.2$, with $u_{+}=0.035$ (thus $\Delta=0.965$). The turning points of these two admissibility criteria, associated with the values of the nonlinearity parameter $\epsilon$ and the initial jump $\Delta$, are depicted in Figures \ref{f:instability}(a) and \ref{f:instability}(a). 
In Fig.~\ref{f:instability}(b), features of modulation instability in the structure of the bore are evident, consistent with the prediction of the associated admissibility condition. Similarly, Fig.~\ref{f:lindeg}(b) reveals non-standard, non-uniform waves evolution at the trailing edge of the bore. In this case, the associated Whitham modulation system does not form a strictly hyperbolic system due to the generation of a multi-speed wavetrain, resulting in the loss of genuine nonlinearity.

\begin{figure}[!ht]
    \centering
    \includegraphics[width=1.0\textwidth]{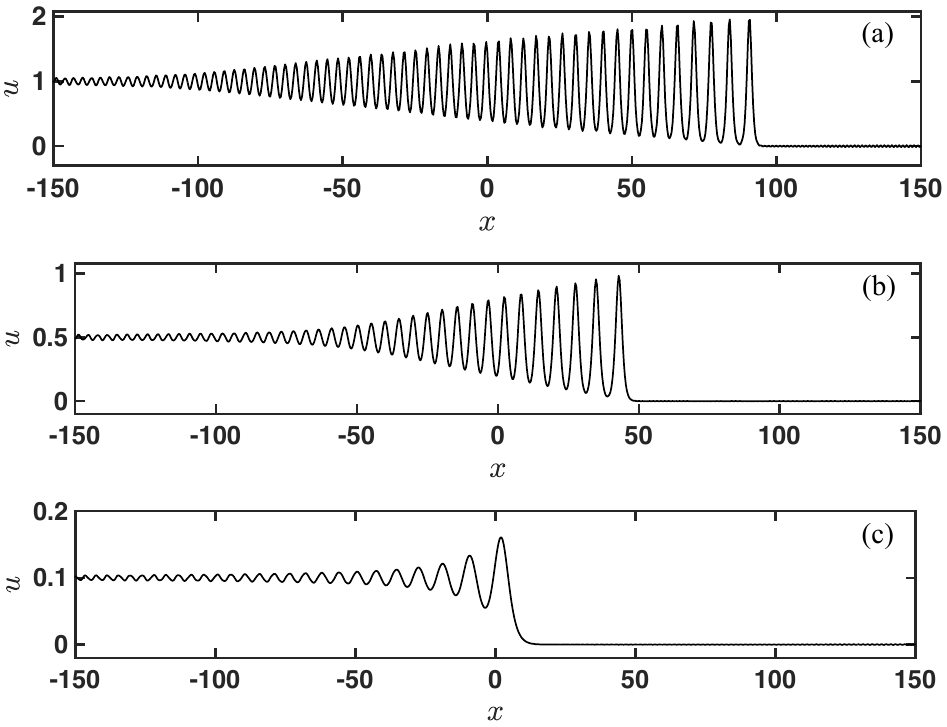}
    \caption{Fulfillment of admissibility conditions, as well as stability and nonlinearity robustness of shallow water undular bores across a broad range of initial discontinuous jumps $\Delta$. (a) Bore at $\Delta=1.0$, (b) bore at $\Delta=0.5$, (c) bore at $\Delta=0.1$. Here, $t=25$, $\epsilon=0.15$, $c_{1}=-3/2$, $c_{2}=23/4$, $c_{3}=5/2$, $c_{4}=19/40$. }
     \label{f:waterwaves}
\end{figure}

Finally, it has been shown \cite{resekdv} that the shallow water wave coefficients effectively suppress the growth of resonant radiation attached to the DSW, thereby stabilizing the DSW's structure as much as possible. However, this observation was based on numerical solutions and lacked theoretical verification. The DSW admissibility condition can provide justification for this phenomenon. Indeed, by setting $\epsilon=0.15$ and utilizing the shallow water coefficients (\ref{e:shallowcoef}), Fig.~\ref{f:waterwaves} illustrates shallow water undular bores across a wide range of initial jumps $\Delta$, exhibiting strong stability and nonlinearity in their structures.

\section{Conclusions and future directions}

In conclusion, in this work we have studied shallow water dispersive shock waves 
(DSWs), or undular bores, in the context of non-convex dispersive hydrodynamics. 
The latter is an emerging branch of the theory of nonlinear dispersive waves, focusing 
on the investigation of non-classical wavetrains, such as solitary waves and DSWs 
in systems exhibiting non-convex dispersion. Here, our system of interest was the 
extended Korteweg-de Vries (eKdV) equation, which results in shallow water wave 
theory one order beyond the usual KdV equation, and incorporates additional higher 
order dispersive and nonlinear terms. As such, the eKdV can model higher amplitude 
and steeper waves. 

The higher order corrections present in the eKdV equation are crucial for the correct 
description of DSWs. In particular, while the dispersion relation of the KdV 
equation is convex, so that resonance between a DSW and dispersive radiation is not 
possible, the higher order terms of the eKdV equation lead to a non-convex 
linear dispersion relation and, hence, resonance between an undular bore and linear 
dispersive waves is possible; this results in the emergence of a resonant wave train 
ahead of the bore. 
We have identified relevant DSW regimes, depending 
on the magnitude of the initial jump that forms the bore and the values of the 
coefficients of the eKdV equation. These regimes, which were depicted by means 
of direct numerical simulations, include   
radiating DSWs (RDSWs) ---relevant to the KdV model--- as well as non-covex 
regimes, namely cross-over DSWs (CDSWs) and traveling DSWs (TDSWs) 
---relevant to the eKdV model.   

To better understand the emergence of the above mentioned DSW regimes, we presented 
an overview of the mathematical methods that are used for the description of DSWs. 
First we focused on the RDSW regime, where bores are stable. To study this case, 
we started by presenting the Whitham modulation theory, relying either on the 
averaging of conservation laws or on the averaging of Lagrangians. This approach 
leads to DSW solutions, typically in cases where the underlying model is integrable 
(such as the KdV equation) and the modulation equations can be put in a Riemann 
invariant form. In non-integrable settings, KdV-type DSWs can be found by El's 
shock fitting method, which relies solely on the knowledge of the linear dispersion 
relation. We have presented this technique using, as an example, the KdV model. 
Furthermore, we examined the admissibility conditions, which are necessary to 
maintain the stable form of eKdV shallow water dispersive shocks. In addition, we 
have also discussed equal amplitude approximation, introduced by Marchant and Smyth, 
an approach relevant to situations where the underlying dispersive hydrodynamic 
system is elliptic in the dispersionless limit; in such a case, the DSW is unstable. 
This method provides important information, leading to the determination of the 
amplitude and velocity of the leading solitary wave within the unstable part of 
the bore in the CDSW regime. 

Next, we turned our attention to the analysis of the resonant radiation in the 
CDSW and TDSW regimes. We thus introduced Whitham shocks, first introduced by Sprenger 
and Hoefer, for the eKdV CDSW and TDSW. A Whitham shock is a moving discontinuity 
connecting the resonant wavetrain ahead with the wavetrain behind, and can be viewed 
as the dispersive equivalent of the Rankine-Hugoniot jump conditions arising in the 
study of classical shock waves (e.g., in the context of gas dynamics). We have found 
that the eKdV TDSW regime is a special case of the regime of eKdV CDSW, occurring when 
the amplitude of the lead solitary wave of the DSW diminishes. 
Results of direct numerical simulations, in the framework of the KdV and eKdV models,  
corroborated the analytical predictions, and the agreement between the two was found 
to be very good.    

Having presented fundamental principles and characteristics of DSWs, 
having explored various analytical approaches for their study, and 
having examined recent advancements in the field of non-convex dispersive 
hydrodynamics, we hope that this work will inspire relevant theoretical studies in 
this direction. There are many interesting themes that remain to be explored, 
such as the study of solitary waves and DSWs of other extended shallow water wave 
equations featuring non-convex dispersion. A relevant investigation also concerns 
the asymptotic integrability of these other higher order weakly nonlinear dispersive 
wave equations. Furthermore, it would be interesting to extend relevant studies in 
quasi-1D (radially-symmetric) and fully 2D settings. Finally, the application of the 
presented methodologies to relevant problems in plasmas, nonlinear optics, and other 
application areas, would be particularly relevant. 

\section*{Acknowledgments}
This work, initiated by discussions and collaboration with Noel F. Smyth, 
and inspired by his major contributions to the field of non-convex dispersive hydrodynamics, is dedicated to his memory \cite{noel}.

\end{document}